\documentclass[fleqn,usenatbib]{mnras}

\usepackage[utf8]{inputenc}
\usepackage{threeparttable}
\usepackage{amsmath}
\usepackage{graphicx}
\usepackage{epstopdf,color,verbatim}
\usepackage{threeparttable}
\usepackage{graphicx, color}
\usepackage{bm}
\usepackage{soul}
\usepackage{amssymb}
\usepackage{hyperref}

\newcommand{\eqb}{\begin{eqnarray}}
\newcommand{\eqe}{\end{eqnarray}}
\newcommand{\bacc}{\beta_{\rm a}}
\newcommand{\bg}{\beta_{\rm g}}
\newcommand{\lacc}{\ell_0}
\newcommand{\fsup}{f_{\rm sup}}
\newcommand{\wmx}{w_{p=\sqrt{\sigma}}^{\max}}
\newcommand{\wbr}{w_{p\rightarrow \sqrt{\sigma}}^{\max}}
\newcommand{\comp}{\,c/\omega_{\rm p}}
\newcommand{\mel}{m_{\rm e}}
\newcommand{\Dtdb}{\Delta t_{1/2}}

\title[Statistics of the plasmoid chain]
{Plasmoid statistics in relativistic magnetic reconnection}

\author[]{M. Petropoulou$^1$, I. M.~Christie$^1$, L. Sironi$^2$, D. Giannios$^1$ \\
$^1$Department of Physics, Purdue University, 525 Northwestern Avenue, West Lafayette, IN, 47907, USA \\
$^2$Department of Astronomy, Columbia University, 550 W 120th St, New York, NY 10027, USA
}

\begin{document}

\date{Received.../Accepted...}

\pagerange{\pageref{firstpage}--\pageref{lastpage}} \pubyear{2017}

\maketitle

\label{firstpage}

\begin{abstract}
Plasmoids, quasi-spherical regions of plasma containing magnetic fields and high-energy particles, are a self-consistent by-product of the reconnection process in the relativistic regime.  Recent two-dimensional particle-in-cell (PIC) simulations have shown that plasmoids can undergo a variety of processes (e.g. mergers, bulk acceleration, growth, and advection) within the reconnection layer. We developed a Monte Carlo (MC) code, benchmarked with the recent PIC  simulations, to examine the effects of these processes on the steady-state size and momentum distributions of the plasmoid chain. The differential plasmoid size distribution is shown to be a power law, $N(w)\propto w^{-\chi}$,  ranging from a few plasma skin depths to $\sim 0.1$ of the reconnection layer's length. We demonstrate numerically and analytically that the power law slope $\chi$ is linearly dependent upon the ratio of the plasmoid acceleration and growth rates and that it slightly decreases (from $\sim 2$ to $\sim 1.3$) with increasing plasma magnetization (from 3 to 50). We perform a detailed comparison of our results with those of recent PIC simulations and briefly discuss the astrophysical implications of our findings through the representative case of flaring events from blazar jets. 
\end{abstract}


\section{Introduction}
Magnetic fields are ubiquitous in a variety of astrophysical sources. How their energy is transferred to the plasma to be later radiated away and power the observed emission remains a fundamental question  in modern high-energy astrophysics. The topological rearrangement of magnetic field lines with opposite polarity -- the so called magnetic reconnection process -- is a mechanism that can transfer the magnetic energy to plasma resulting in heating  and particle acceleration.

The reconnection process has been studied extensively and is an excellent candidate for producing variable, high-energy radiation in many astrophysical environments such as pulsar wind nebulae  \citep[PWNe; e.g.,][]{lyubarsky_kirk_01,lyubarsky_03,kirk_sk_03,petri_lyubarsky_07,sironi_spitkovsky_11b, cerutti_13a,cerutti_14, lyutikov_16}, jets from active galactic nuclei \citep[AGNs; e.g.,][]{romanova_92,giannios_09,giannios_10b,giannios_13}, gamma-ray bursts \citep[GRBs; e.g.,][]{thompson_94, thompson_06,usov_94,spruit_01,drenkhahn_02a,lyutikov_03,giannios_08}, accreting black holes \citep{galeev_79, haardt_91, belo_17}, and solar flares \citep{lin_05, hassanin_16}.
 
Magnetic reconnection is a highly dynamical process for the plasma conditions present in the aforementioned  astrophysical environments.
The reconnection layer is prone to tearing instabilities \citep[e.g.][]{furth_63, loureiro_07, uzdensky_16, comisso_16} and can undergo fragmentation into magnetic islands, the so-called plasmoids. These are, in turn, separated by secondary current sheets that are shorter and still subject to the same instabilities. This fragmentation process recurs at progressively smaller scales that exhibit an approximate self-similarity in their properties \citep{shibata_01, huang_10}. The fractal-like appearance of the reconnection layer may also be related to the generation of power laws \citep{schroeder_91} that allow the extrapolation and prediction over a wide range of scales. 
 
Recent analytical calculations and magnetohydronamic simulations of  magnetic reconnection in highly conducting plasmas
have been used to determine the statistical properties of the resulting plasmoid chain. The plasmoid magnetic flux
distribution was shown to obey a power-law of slope $-2$ \citep{uzdensky_10, Loureiro_12} or $-1$ \citep{huang_12, huang_13}, depending on the underlying assumptions made in each case. 

In the collisionless regime of reconnection, which is applicable to most astrophysical environments, the most fundamental way to capture the interplay between particles and fields is by means of kinetic particle-in-cell (PIC) simulations. PIC simulations of reconnection have been recently extended to 
the so-called relativistic regime, where the magnetic energy density is much larger than the rest-mass energy density of the unreconnected plasma, or the plasma magnetization $\sigma \gtrsim 1$ \citep{zenitani_01, guo_14, ss_14, nalewajko_15, spg_15, kagan_16, werner_16}. This regime is most likely relevant to AGN jets, pulsar winds, and  GRBs.  \citet{sironi_16} (hereafter, SGP16) performed two-dimensional (2D) PIC simulations of antiparallel reconnection (i.e., in the absence of a guide field) in electron-positron plasmas for three different plasma magnetizations. The simulations revealed 
the rich and complex dynamics of the reconnection layer. Being extended to unprecedentedly long time scales and length scales, they also allowed the investigation of the properties of the plasmoid chain (e.g., the geometry, the momentum, the particle and magnetic energy content of individual plasmoids) as a function of time and system size. 

The aim of this study is to shed light into the processes that shape the plasmoid size and momentum distributions in light of 
recent PIC simulations of relativistic magnetic reconnection. To achieve our goal, we develop a Monte Carlo (MC) code that follows the evolution of individual plasmoids as they grow, accelerate, merge or leave the layer. We calibrate the MC code using some findings of the  SGP16 PIC simulations (e.g., plasmoid acceleration rate), but we also make predictions (e.g., distribution of plasmoid four-velocities) that compare well with PIC results.

Although the MC approach cannot replace first-principle PIC simulations of the reconnection layer (e.g., it carries no information on particle acceleration), it offers several advantages over them:
\begin{enumerate}
 \item it is computationally cheaper, thus allowing the realization of many numerical experiments for the same physical conditions of the system (e.g., plasma magnetization). The repetition of numerical experiments is particularly important in studies focusing on the statistical properties of a system (here, the plasmoid chain) whose evolution is governed by some intrinsically random process (e.g., mergers). 
 \item it facilitates the study of the system over long temporal and spatial scales without the need of extensive computational resources. In fact, the extreme separation between the microscopic plasma scales that PIC simulations need to resolve and the large physical scales where the emission takes place (e.g., nine orders of magnitude in AGN jets) usually hinders a direct application of PIC findings to astrophysical observations.
 \item it allows us to isolate the role of individual physical processes, such as plasmoid acceleration and mergers, on the statistical properties of the plasmoid chain. This is achieved by artificially turning off individual processes in the MC code. This approach is not possible in PIC simulations.
 \end{enumerate}
 
Using the MC code we have developed, we find that the differential plasmoid size distribution forms a power law, $N(w)\propto w^{-\chi}$, 
beyond a characteristic plasmoid size of several plasma skin depths\footnote{The electron skin depth is defined as $\comp \equiv  c \sqrt{\mel/(4\pi n e^2)}$, where $n$ is the electron number density of the unreconnected plasma and $e$ is the electron's charge.}. We find that the power-law slope $\chi$ decreases from $\sim 2$ to $\sim 1.3$ as the magnetization $\sigma$ increases from 3 to 50. The slope of the power law is proportional to the ratio of the plasmoid acceleration rate to the plasmoid growth rate ($\chi \propto \bacc/\bg$), a result we also derive analytically, and is not altered by mergers. We show that the formation of a power law in sizes is the result of the interplay between plasmoid acceleration and growth. The power law extends from small sizes (i.e., several plasma skin depths) to large sizes that are a significant fraction  (i.e., up to $\sim$ 10 percent) of the reconnection layer's length. As the system becomes larger, so does the extent of the power law. The cutoff of the distribution at large sizes is populated by a few plasmoids or ``monster'' plasmoids \citep{uzdensky_10}. The cutoff shifts towards larger sizes almost linearly with time until the monster plasmoids exit the system. 
We also measure the size distribution of plasmoids at the moment they exit the layer and find that it is similar to the position-integrated
distribution of sizes, as expected in a steady state system. We demonstrate that the differential distribution of plasmoid momenta becomes softer at higher $\sigma$, in agreement with PIC results, and does not depend on the system's size.  Our main results can also be extended to high magnetization ($\sigma \gtrsim 3$) electron-proton reconnection, since the system behaves similarly to the electron-positron case \citep{spg_15}.
 
This paper is structured as follows. In Sect.~\ref{sec:pic} we summarize the basic findings of the PIC simulations presented in SGP16. These were used in the development of the MC code, which is outlined in Sect.~\ref{sec:method}.  The results of our MC simulations of the reconnection layer are presented in Sect.~\ref{sec:results}. An application of our results to blazar variability is presented in Sect.~\ref{sec:astro}. 
We conclude in Sect.~\ref{sec:summary} with a short discussion and summary of our results. 

\section{Summary of PIC results}\label{sec:pic}
SGP16 employed large-scale 2D PIC simulations in electron-positron plasmas for three different magnetizations ($\sigma=3,10$, and 50). The 
simulations were extended to unprecedentedly large spatial scales (i.e., thousands of electron skin depths). SGP16 were able to follow the system evolution for several light crossing times of the layer, due to the use of outflow boundary conditions instead of the commonly employed periodic boundary conditions. The simulations were able to capture the dynamics of the reconnection layer at times when the system was no longer affected by the initial setup. SGP16 showed that plasmoids are continuously generated as a self-consistent by-product of the reconnection process and highlighted the dynamic nature of the layer. 

Here, we summarize the main findings of PIC simulations that are relevant to the calculations presented in this paper (see Sect.~\ref{sec:method}) and we refer the reader to SGP16 for more details.
\subsection{Plasmoid birth} 
\begin{figure}
 \centering 
  \includegraphics[width=0.47\textwidth]{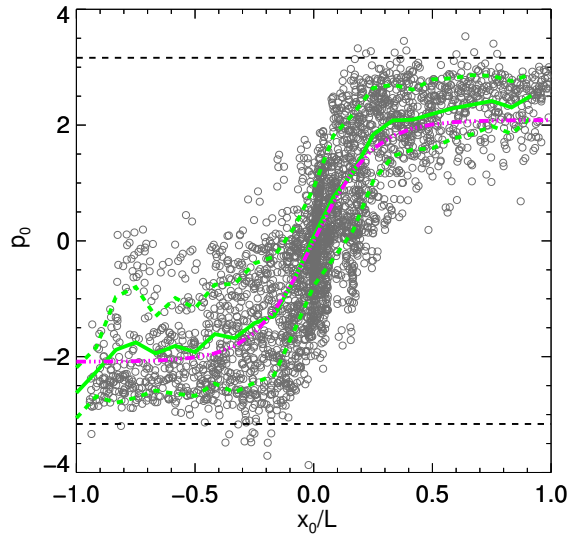}
 \caption{Plot of the initial plasmoid momentum versus the initial plasmoid location in the layer for plasma magnetization $\sigma=10$ (open symbols). All plasmoids formed in the layer during the course of the PIC simulation are taken into account. Data are obtained from SGP16. The horizontal black dashed lines indicate the asymptotic momentum $\sqrt{\sigma}$. The average initial four-velocity of the plasmoid distribution (solid green line) can be well described by eqn.~(\ref{eq:p0}) shown by the magenta dash-dotted line. Dashed green lines indicate the one standard deviation of $p_0$.}
 \label{fig:init-dist}
\end{figure}

Plasmoids are constantly generated due to the fragmentation of secondary current sheets. The initial plasmoid size is
a few electron skin depths and, thus, a small fraction of the layer's length. The separation of neighboring plasmoids at birth is on average ten times larger than their initial size. Plasmoids are uniformly formed in the available free spaces of the layer at any time. We indicate the plasmoid location at birth with $x_0$. The separation of plasmoids born close to the center of the layer and on opposite sides with respect to it quickly increases, as the tension force of the field lines drags them in opposite directions\footnote{In the PIC simulations of SGP16, reconnection is triggered at the center of the layer, a choice that eventually determines the geometry of the field lines in steady state, see Fig.~1 in SGP16.}. Thus, as time progresses the available locations for additional plasmoid generation become concentrated close to the center of the layer. This results in a  distribution of birth locations that peaks at $x_0\approx 0$ at later times. 
The initial plasmoid momentum\footnote{Henceforth, we use the term momentum and dimensionless four-velocity interchangeably.}, $p_0$, is found to correlate with the birth location. This is exemplified  in Fig.~\ref{fig:init-dist}, where $p_0$ is plotted against $x_0$ for all the plasmoids formed in the layer in the course of the PIC simulation with $\sigma=10$. The average $p_0$ of the plasmoid population is described well by the relation
 \eqb 
 \langle p_0 \rangle = 0.66 \sqrt{\sigma} \tanh \left( \frac{x_0}{\lacc} \right),
 \label{eq:p0}
  \eqe 
where $\sqrt{\sigma}$ is the asymptotic bulk four-velocity  in the reconnection layer \citep{lyubarsky_05}. In fact, the distribution of the differences $\langle p_0 \rangle-p_{0}$ (residuals) is approximately Gaussian with zero mean and standard deviation of $0.3\sqrt{\sigma}$ \citep{andrae_10}. The characteristic length scale, $\lacc$, also depends on the plasma magnetization as:
 \eqb 
 \lacc \approx 0.25\, L \left(\frac{\sigma}{10}\right),
 \label{eq:accscale}
\eqe 
where $L$ is the half length of the layer.  This implies that at higher magnetizations few plasmoids 
are born with momenta as large as $\sqrt{\sigma}$. As a consequence, few plasmoids will reach the terminal momentum $\sqrt{\sigma}$ in the course of their evolution (see also SGP16). 

\subsection{Plasmoid growth} Secondary plasmoids grow mostly via the accretion of smaller plasmoids. SGP16 demonstrated that during the most active phase of growth the plasmoid size increases as:
\eqb 
\frac{{\rm d}w}{{\rm d}t}=\frac{\bg c}{\gamma}
\label{eq:growth}
\eqe 
where time is measured in the layer's frame, $\gamma=\sqrt{1+p^2}$ is the Lorentz factor of the plasmoid and $\bg$ is the growth rate as measured in the co-moving frame of the plasmoid. SGP16 demonstrated that $\bg$ is about half of the reconnection inflow rate which, in turn, depends only weakly on $\sigma$ (see Table~\ref{tab:param}). Although the equation  above provides an overall good description of the average plasmoid growth, it over-predicts the growth of the fastest plasmoids -- see Fig.~8 in SGP16 and  Appendix~\ref{app:app0} for more details. We therefore introduce a ``suppression factor'' of the  plasmoid growth, which should multiply the right-hand side of eqn.~(\ref{eq:growth}). This depends upon the plasmoid four-velocity $p$ and is given by:
\eqb 
\fsup(p) = \left[\frac{1}{2}\left(1 - \tanh \left( \frac{\frac{|p|}{\sqrt{\sigma}}- A}{B} \right) \right) \right]^{-1}
\label{eq:sup}
\eqe 
where $A,B$ are parameters to be determined (see Appendix~\ref{app:app0}). For the values of $\sigma$ explored here, $A/B \gg 1$ (see Table~\ref{tab:param}), so that $\fsup \rightarrow 1$ when $|p|/\sqrt{\sigma}\ll 1$ (i.e., no suppression of the growth at non-relativistic speeds). 
\subsection{Plasmoid acceleration} Secondary plasmoids are accelerated to the Alfv\'en speed ($v_{\rm A}/c=\sqrt{\sigma/(1+\sigma)}$) by the tension force of the reconnected magnetic field. The momentum $p$ of an individual plasmoid at time $t$ is found to depend upon the position of its center in the layer $x$ and its size $w$ as (see Fig.~10 in SGP16):
 \eqb
p(t) \approx \sqrt{\sigma} \tanh\left(\frac{\bacc}{\sqrt{\sigma}}\frac{x(t)-x_0}{w(t)}\right) + p_0,
\label{eq:p}
\eqe
where $t$ is measured in the layer's frame and $\bacc$ is the acceleration rate of the plasmoid, which is only weakly dependent upon $\sigma$ (see Table~\ref{tab:param}).

\section{Monte Carlo code} \label{sec:method}
Here, we present the main ingredients of the MC code and its calibration using the SGP16 simulation results. 

All plasmoids are born with transverse size of several $\comp$, which is typically a small fraction of the layer's half-length $L$. For most of our results and for their direct comparison with the PIC simulations (Sect.~\ref{sec:results}), we set the plasmoid size at birth  $w_*=10^{-3}\,L$.  In Sect.~\ref{sec:size} we explore how our results change for smaller values of $w_*/L$, effectively corresponding to larger system sizes. The transverse size refers to the width of the plasmoid in the direction perpendicular to its motion and is Lorentz invariant. The longitudinal size as measured in the co-moving frame of the plasmoid is  $w_{\parallel} \simeq (3/2) w$ at all times (see Fig.~5 in SGP16). 

The separation distance between two newly born neighboring plasmoids is $\delta x \sim 10 w_*$, as found in PIC simulations. At the beginning of the MC simulation, the total number of plasmoids is, therefore, $N_0=2 L/\delta x$.  At later times, new plasmoids are formed in regions of the layer where there is unoccupied space. The new plasmoids are born at random locations $x_0$ within the available free space, with the same size and separation distance as the initial ones.

The plasmoid momentum at birth is drawn from a Gaussian distribution of random numbers. The mean of the distribution is given by eqn.~(\ref{eq:p0}) and its standard deviation is $0.3\sqrt{\sigma}$. Although the number of plasmoids with $|p_0| > \sqrt{\sigma}$ is small (i.e., a few percent of the total number), we impose   a hard upper limit of $1.5 \sqrt{\sigma}$ on the initial four-velocity. Any plasmoid that initially does not satisfy this condition is rejected and another random $p_0$ value is drawn from the distribution.    

\begin{figure}
 \centering 
\includegraphics[width=0.47\textwidth]{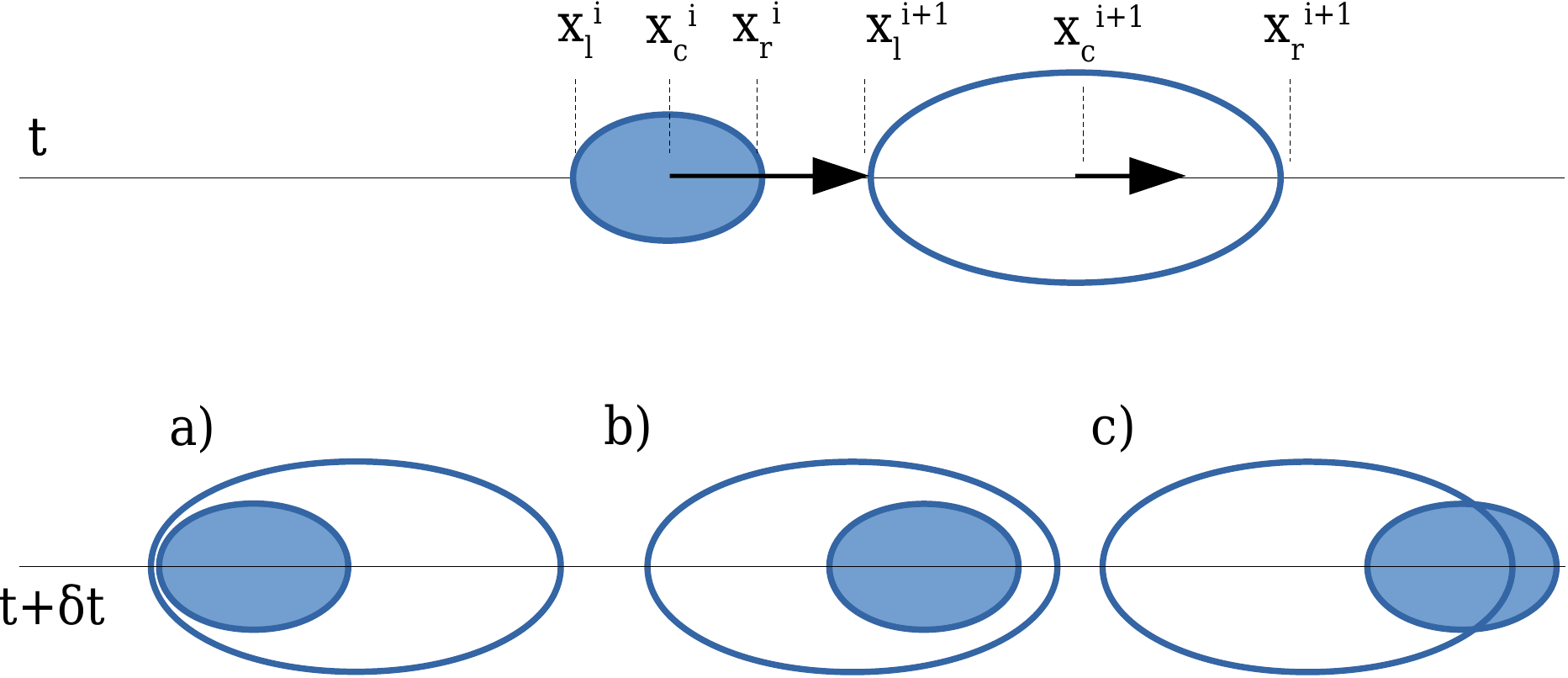} 
\caption{Sketch illustrating different merging scenarios of two neighboring plasmoids moving towards the same direction (to the right, in this case).} 
\label{fig:sketch}
\end{figure} 

At every time step, we identify the plasmoids that terminate their lives due to advection beyond the layer by checking if their inner boundary 
is located outside the layer at time $t$, namely:
\eqb 
x_{l}(t) \equiv x(t)-\frac{1}{2}\frac{w_{\parallel}(t)}{\sqrt{1+p(t)^2}} > L, \,  x(t) > 0 \\
x_{r}(t) \equiv x(t)+\frac{1}{2}\frac{w_{\parallel}(t)}{\sqrt{1+p(t)^2}} < -L, \,  x(t) < 0,
\eqe 
where $x$ is the location of the plasmoid's center at time $t$.
Mergers are another way of terminating the life of a plasmoid. At every time step, we check for mergers between existing plasmoids. 
As in Fig.~\ref{fig:sketch}, let us consider the case  $x_c^i(t)<x_c^{i+1}(t)$ and assume that the two plasmoids move to the right.
A merger is defined based on one of the following criteria: 
\begin{enumerate}
 \item $x_{r}^{i}(t+\delta t) > x_{r}^{i+1}(t+\delta t)$. This suggests that  an intersection of the outer boundaries of the two plasmoids took place.
 \item $x_{c}^{i}(t+\delta t) > x_{c}^{i+1}(t+\delta t)$.  This suggests that an intersection of the plasmoids' centers took place.
 \item $x_{r}^{i}(t+\delta t) > x_{r}^{i+1}(t+\delta t)$ or $x_{l}^{i}(t+\delta t) > x_{l}^{i+1}(t+\delta t)$. This suggests that an intersection of the two plasmoids took place.
\end{enumerate}
Case (a) shown in Fig.~\ref{fig:sketch} would be identified as a merger only using criterion (iii), whereas case (b) would be recognized as a merger by criteria (ii) and (iii). All the above criteria would identify case (c) as merger. For the results presented in the next section, we use criterion (i), but we demonstrate how different criteria affect our results in Sect.~\ref{sec:mergers}.  The smaller plasmoid of the merging pair is removed from the simulation after the merger. The plasmoid surviving the merger is assumed to have the same size as the larger plasmoid of the merging pair, since the average growth rate $\bg$ already includes the effect of mergers.

At every time step, we identify all the plasmoids that are still present in the layer and update their transverse size and momentum. We first advance the position of a plasmoid according to $x(t) \rightarrow x(t)+ v(t) \delta t$, where $v$ is the plasmoid velocity and $\delta t$ is the time step of integration. In order to capture the evolution of the fastest plasmoids in the layer,  we use $\delta t < \delta x / v_A$. The transverse size is then updated according to eqn.~(\ref{eq:growth}) including the suppression factor. Finally, the plasmoid momentum is updated according to eqn.~(\ref{eq:p}). 

\begin{figure}
 \centering
\includegraphics[width=0.47\textwidth]{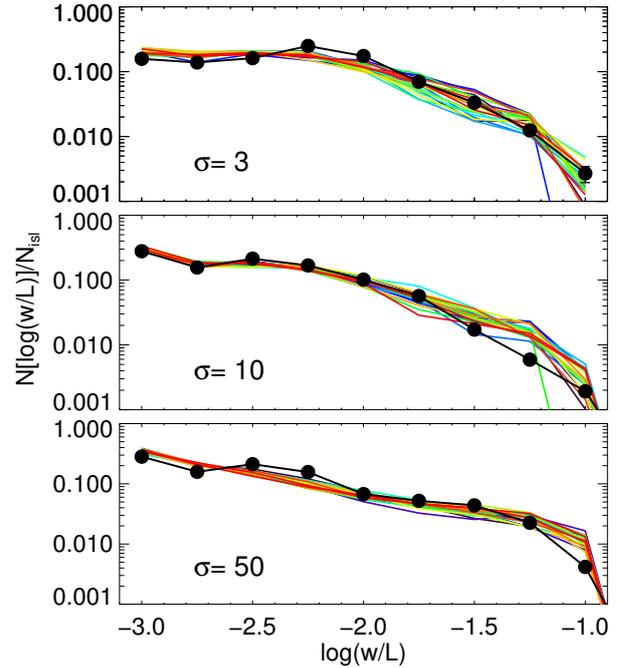}
\caption{Histograms of $\log(w/L)$ normalized to the total number of plasmoids $N_{\rm isl}$ for the three magnetizations considered in the text. The distributions obtained from the PIC simulations of SGP16 are shown with filled black points connected by black lines. Coloured lines are the distributions from 20 MC realizations. }
\label{fig:comparison-1}
\end{figure} 

\begin{figure}
 \centering
\includegraphics[width=0.47\textwidth]{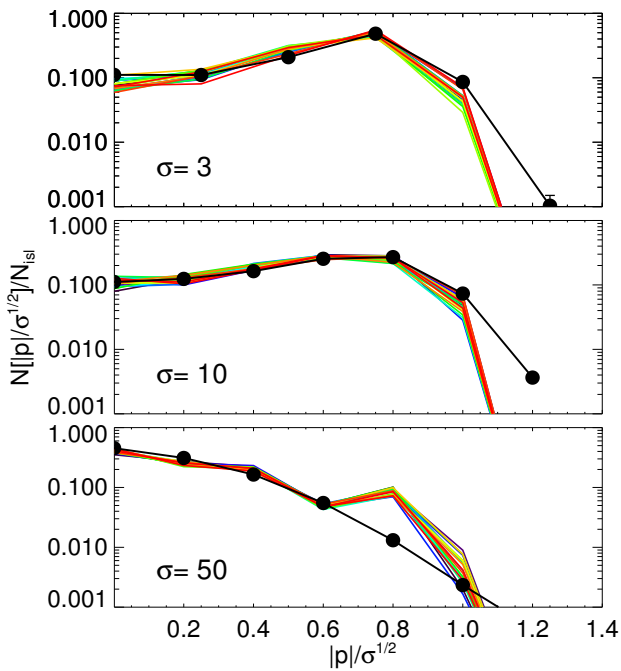}
\caption{Histograms of $|p|/\sqrt{\sigma}$  normalized to the total number of plasmoids $N_{\rm isl}$ for the three magnetizations considered in the text. Symbols and lines have the same meaning as in Fig.~\ref{fig:comparison-1}.}
\label{fig:comparison-2}
\end{figure}  

\section{Results}\label{sec:results}
In this section, we compare the results of the MC code described in Sect.~\ref{sec:method} against PIC simulations of relativistic magnetic reconnection with respect to the statistics of the plasmoid chain. We show that the basic properties of the plasmoid chain seen in detailed PIC simulations of the reconnection layer can be recovered by the MC code. We then compare our results with those of previous studies of the plasmoid chain statistics \citep{uzdensky_10, huang_12}. Finally, we study the role of individual processes, such as plasmoid acceleration and merging, on the shape of the plasmoid size distribution. 
\subsection{Comparison of MC with PIC results}\label{sec:comparison}
The distributions of plasmoid sizes (${\rm d}N/{\rm d}\log w$) and momenta (${\rm d}N/{\rm d}|p|$) are useful diagnostics of the plasmoid chain  statistics. Henceforth, we adopt the notation $N(X) \equiv {\rm d}N/{\rm d}X$ to refer to the differential distribution of plasmoids with respect to quantity $X$. The distributions are position- and time-integrated, unless stated otherwise. The distribution of magnetic fluxes $\Psi$ is expected to approximately follow the size distribution, since $\Psi \propto w$ for $w \gtrsim 10 w_*$ (see Fig.~5 in SGP16).  The parameters of our MC code have been benchmarked with the PIC simulations of SGP16 and are presented in Table~\ref{tab:param}. 

\begin{table}
\centering
\caption{Parameters of the MC code that are benchmarked with PIC simulations of reconnection.  For the determination of parameters $A$ and $B$, see Appendix~\ref{app:app0}. }
\begin{tabular}{c ccc}
\hline 
$\sigma$ & 3 & 10 & 50 \\
\hline
$\bacc$ & 0.10 & 0.12 & 0.13 \\
$\bg$ & 0.06 & 0.08 & 0.10 \\
$\lacc/L$ & 0.075 & 0.25 & 1.25 \\
$A$& 0.86 &  0.77 &  0.63 \\
$B$& 0.019 &  0.033 &  0.024 \\
\hline
 \end{tabular}
\label{tab:param}
\end{table}  

We simulated the formation of the plasmoid chain using the MC code for $\sigma=3, 10$, and 50.  We let the system evolve for two light crossing times ($2L/c$). This period is comparable to the time interval during which secondary plasmoid generation occurs in PIC simulations along the whole layer
(for details, see SGP16). A comparison of the distributions obtained from our MC code and the PIC simulations of SGP16 is presented in Figs.~\ref{fig:comparison-1}-\ref{fig:comparison-2}. The results from the SGP16 simulations are plotted with black symbols, while coloured lines show the results of 20 MC realizations. The size distributions can be approximated by a broken power law (for $\sigma=3$ and 10) or a single power law (for $\sigma=50$) that cuts off at sizes that are a significant fraction of the layer's length (here, at $w\sim 0.1 \, L$). More details about the shape of the size distribution can be found in Sect.~\ref{sec:breakdown}.

The scatter observed in the MC results stems from the intrinsic randomness of the MC approach and is typically much larger than the errors associated with the number of counts per bin. There is an overall good agreement between the MC results (coloured lines) and PIC results (black symbols) for the plasmoid size distribution (Fig.~\ref{fig:comparison-1}). This is not unexpected, since we used the plasmoid size distribution in order to determine the values of the parameters $A$ and $B$ appearing in the growth suppression factor (for details, see Appendix~\ref{app:app0}). 

Still, it is not obvious {\sl a priori} if the MC code can reproduce the distribution of plasmoid momenta, ${\rm d}N/{\rm d}(|p|/\sqrt{\sigma})$, which is an independent diagnostic. As shown in Fig.~\ref{fig:comparison-2}, there is an overall good agreement between the MC and PIC results. The momentum distribution obtained from PIC simulations extends beyond $\sqrt{\sigma}$, most likely due to the acceleration of a few plasmoids born in the vicinity of larger plasmoids where the local  Alfv{\'e}n  velocity is higher. This effect is not included in our general description for plasmoid momentum in eqn.~(\ref{eq:p}),  which was benchmarked using moderately large plasmoids from PIC simulations. Thus, a deviation between the MC and PIC results is expected for the highest four-velocities. We also find an excess of plasmoids with $|p|>0.7 \sqrt{\sigma}$ with respect to the PIC results for the $\sigma=50$ case. This discrepancy implies that our chosen general prescription for the evolution of the plasmoid momentum with time cannot capture in full detail the plasmoid dynamics for the  $\sigma=50$ case. 

\begin{figure}
 \centering 
\includegraphics[width=0.47\textwidth]{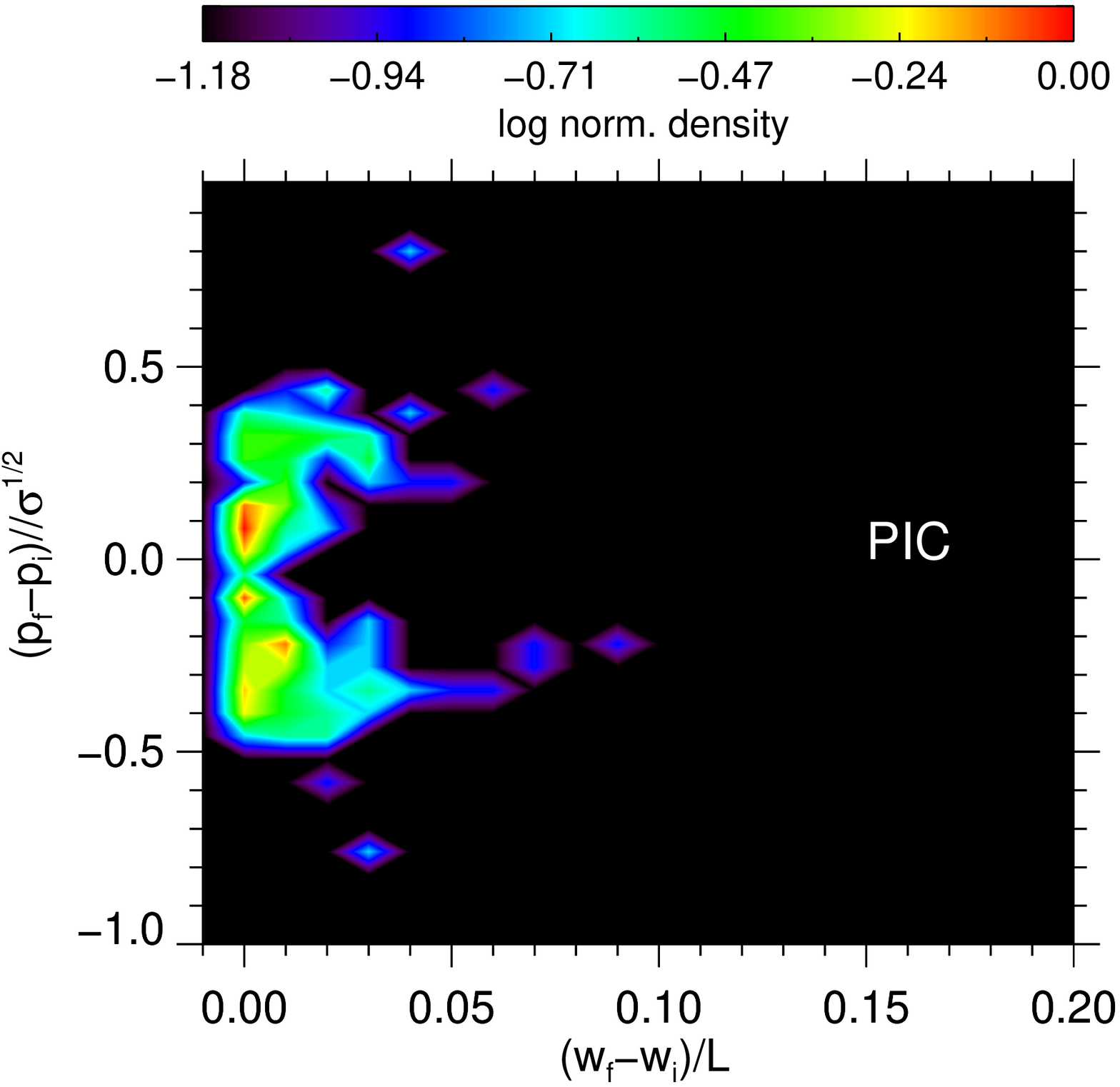} 
\includegraphics[width=0.47\textwidth]{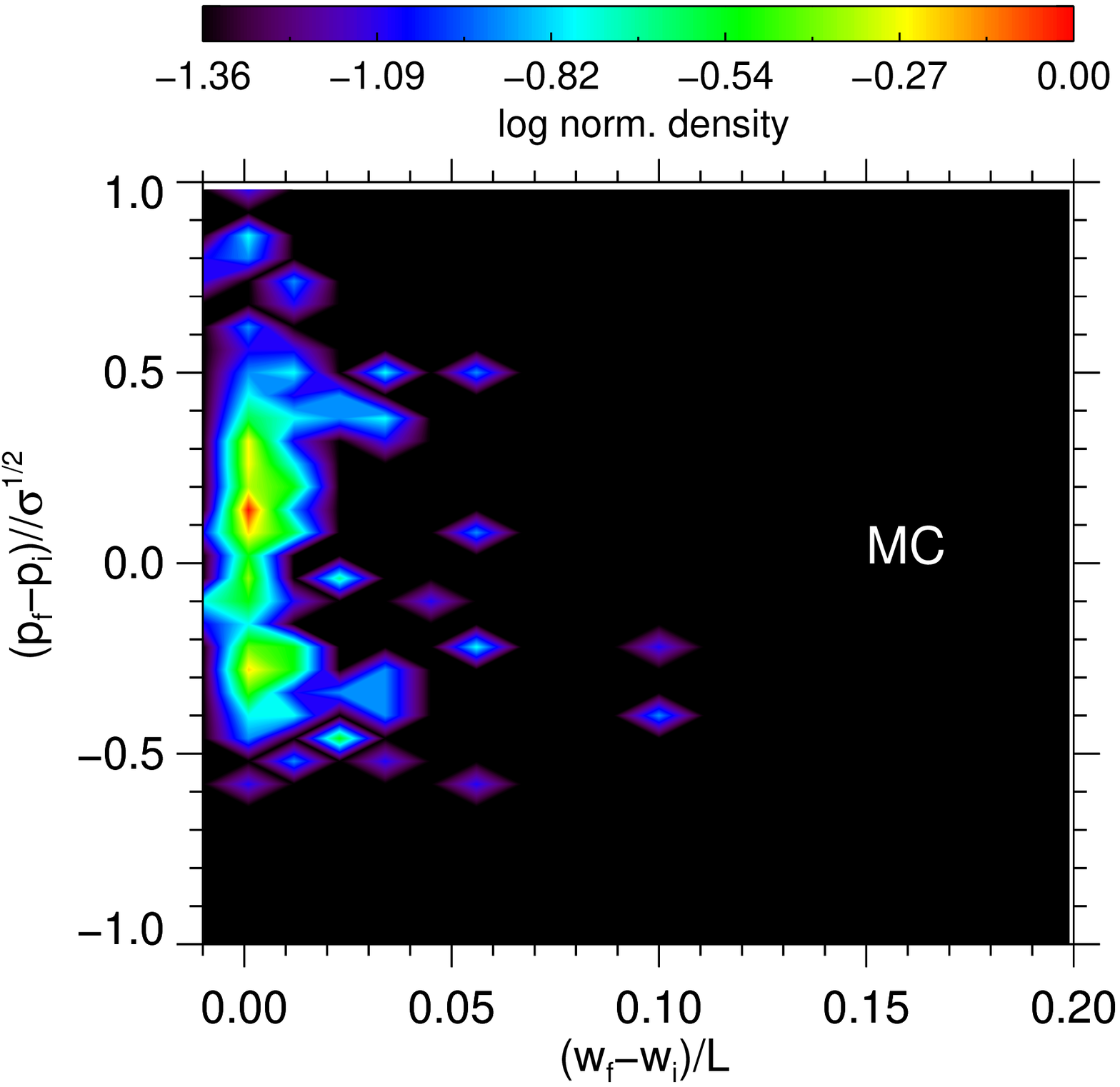} 
\caption{Density map of the relative four-velocity($\Delta p \equiv p_{\rm f}-p_{\rm i}$) and size ($\Delta w \equiv w_{\rm f}-w_{\rm i}$) of  356 merging plasmoid pairs with $w_{\rm i}>0.003\, L$ and $\sigma=10$.  The results of a PIC simulation and a MC realization are shown in the top and bottom panels, respectively.}
\label{fig:merging-1}
\end{figure}

Let us take a closer look at the properties of plasmoids that undergo mergers (see Fig.~\ref{fig:merging-1}). Quantities with subscript ``f'' correspond to the plasmoid that survives the merger. Similarly, the properties of the plasmoid that will be  absorbed by the larger one during the merger are denoted with the subscript ``i''. These properties in the MC code are measured just before a merger. However, in PIC simulations this is not always possible due to the limited time sampling in output data that will be used for the post-processing. For example, two plasmoids (especially those with small sizes) may be born and merge within a time window during which no PIC data have been recorded. These plasmoids will not be counted as a merging pair but as one plasmoid, thus producing a bias towards a smaller number of merging pairs at small sizes. Hence, we consider only plasmoids with $w_{\rm i} > 3 \ w_*$ which had some time to grow and can be confidently identified when post-processing the PIC data.

\begin{figure}
\centering
\includegraphics[width=0.45\textwidth]{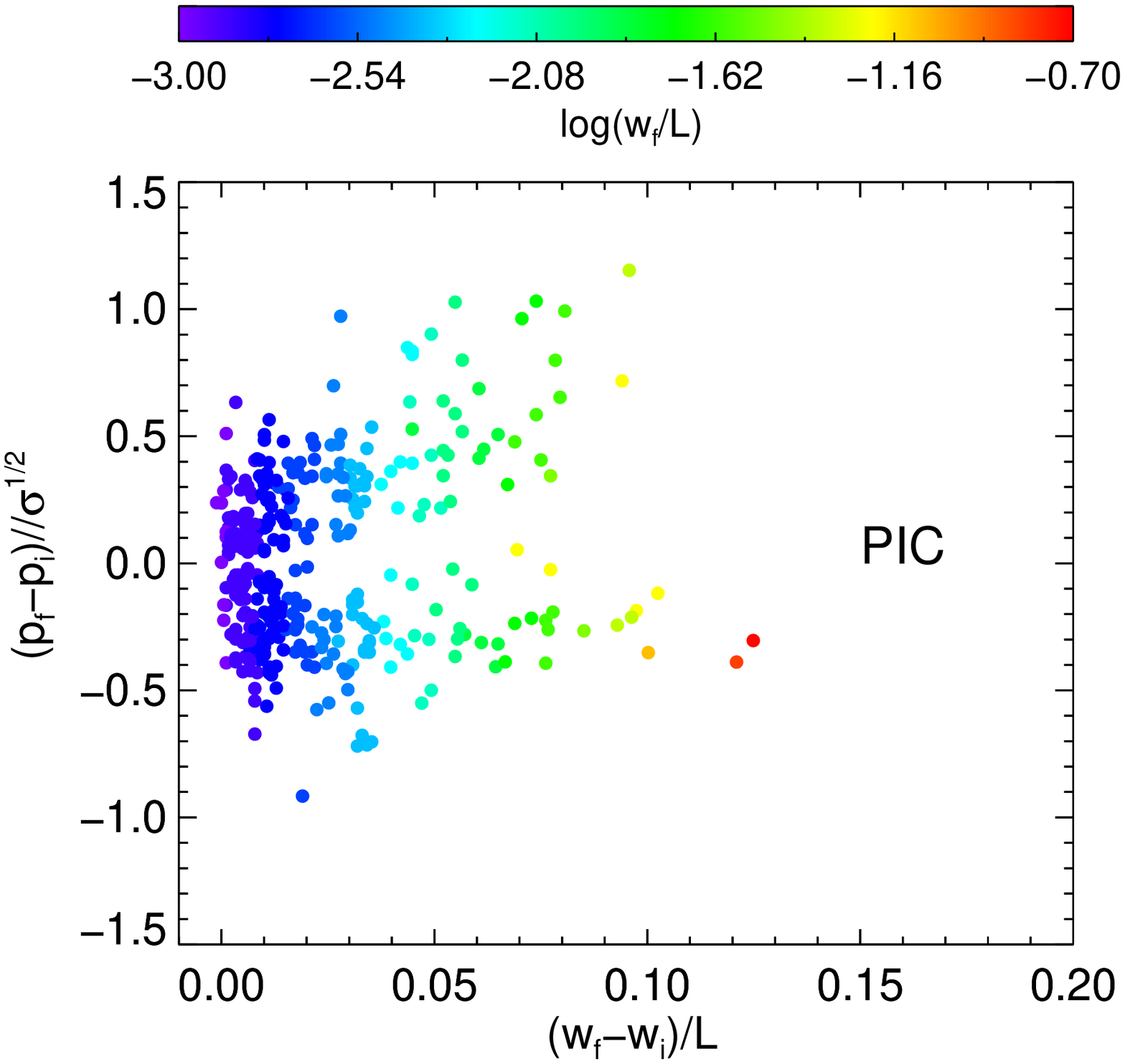} 
\includegraphics[width=0.45\textwidth]{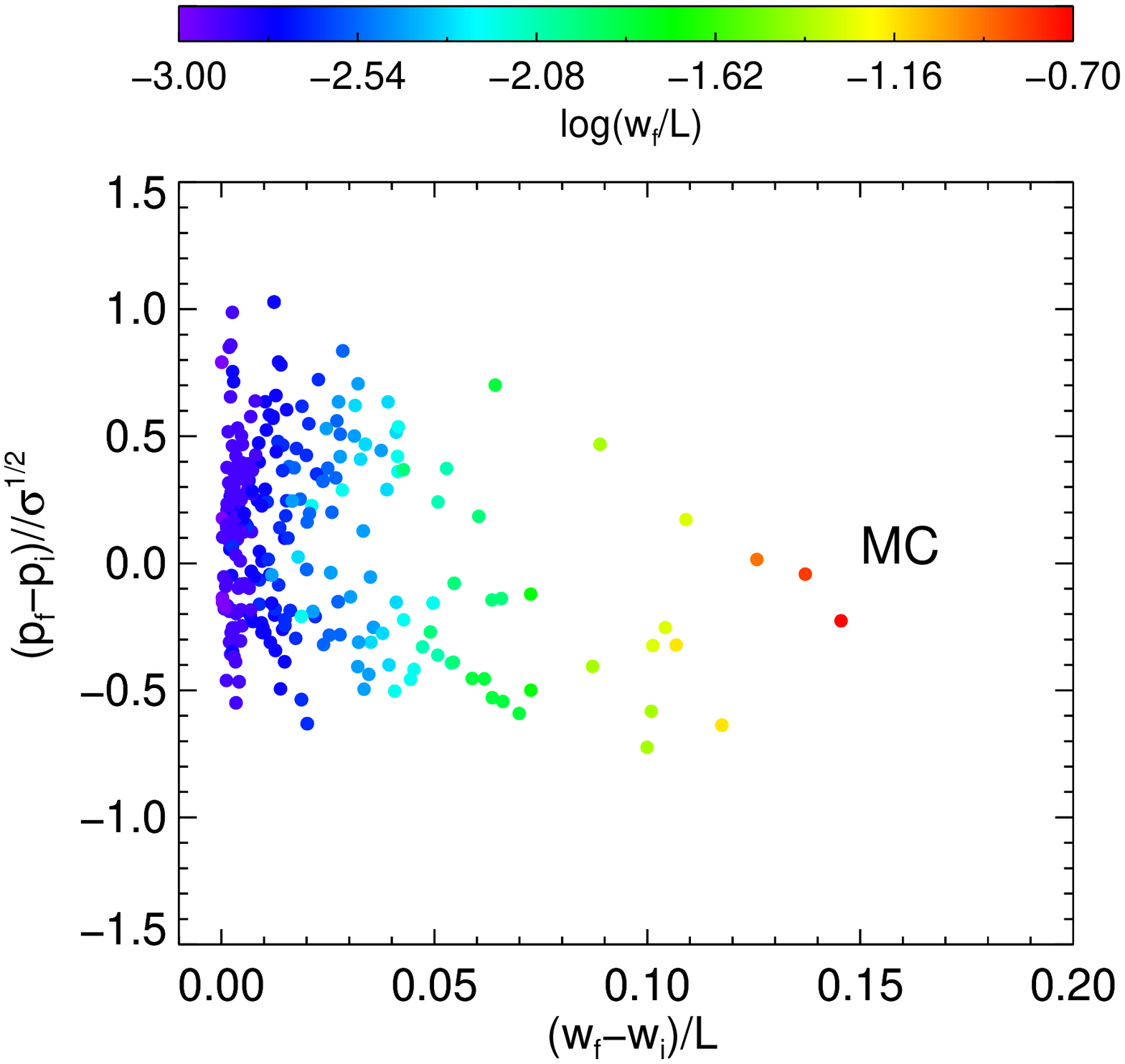} 
\caption{Relative four-velocity ($\Delta p \equiv p_{\rm f}-p_{\rm i}$) and size ($\Delta w \equiv w_{\rm f}-w_{\rm i}$) of 356 merging plasmoid pairs with $w_{\rm i}>0.003 \, L$ and $\sigma=10$. The size of the plasmoid that survives the merger is colour coded, as shown in the colour bar at the top.  The results of a PIC simulation and a MC realization are shown in the top and bottom panels, respectively.}
\label{fig:merging-2}
\end{figure}

Figure~\ref{fig:merging-1} shows a density map of the relative four-velocity and relative size of merging pairs obtained in one of the SGP16 simulations (top panel) and in one of our MC realizations for $\sigma=10$ (bottom panel). To facilitate the comparison of the two, we randomly selected a sub-sample of pairs from the MC code that matches the total number of merging pairs from PIC.  A similar plot is presented in Fig.~\ref{fig:merging-2} with the additional information of the plasmoid size after the merger, $w_{\rm f}$ (see colour bar). A few things that are worth mentioning follow: 
\begin{itemize}
 \item the density of merging pairs in the MC realization peaks at $w_{\rm f}-w_{\rm i} \approx 0$ and $|p_{\rm f} - p_{\rm i}| \lesssim 0.5\sqrt{\sigma}$, in agreement with PIC results (Fig.~\ref{fig:merging-1}). Most of the merging plasmoids have similar sizes of the order of $w_{*}$ (i.e.,  $w_{\rm i} \approx 3 \ w_*$, see Fig.~\ref{fig:merging-2}).
 \item the MC code also predicts a population of plasmoids with large size difference and $|p_{\rm f} - p_{\rm i}|\sim 0.5\sqrt{\sigma}$ at merger. This is in excellent agreement with  PIC results (Figs.~\ref{fig:merging-1}-\ref{fig:merging-2}).
 \item the MC code produces a few merging pairs with $|p_{\rm f} - p_{\rm i}| \approx \sqrt{\sigma}$ and $w_{\rm f}-w_{\rm i}  \approx 0$ that are absent in the PIC simulation. These are plasmoids that did not have time to grow ($w_{\rm f} \approx w_*$) and quickly merged because of their opposite  motions at initialization. Thus, this discrepancy with the PIC results is related to the initialization of the plasmoid momentum in the MC code. The difference would become less prominent, if  a smaller  scatter around $\langle p_0\rangle$ (see eqn.~\ref{eq:p0}) were to be used. 
 \item the differences in the properties of the merging pairs between PIC simulations and MC calculations do not seem to strongly affect  the size and momenta distributions (see Figs.~\ref{fig:comparison-1}-\ref{fig:comparison-2}).  
 \item the distribution of merging pairs in Figs.~\ref{fig:merging-1} and \ref{fig:merging-2} is subject to the randomness of the reconnection process. We verified this by creating the same plots for different MC realizations. This does not alter the aforementioned results, though.
\end{itemize}

\subsection{Comparison with other studies}
The statistics of the plasmoid chain in  magnetic reconnection has been discussed in detail by \citet{uzdensky_10} and \citet{huang_12}. Using heuristic arguments  \citet{uzdensky_10} showed that ${\rm d}N/{\rm d}w \equiv N(w) \propto w^{-2}$ for a range of plasmoid sizes where the merging rate approximately equals  the growth rate. On the contrary, \citet{huang_12} suggested that $N(w) \propto w^{-1}$. The main difference between the two approaches lies in the assumptions made about the relative velocity between merging plasmoids. More specifically, \citet{uzdensky_10} assumed that all plasmoids move with the Alfv{\'e}n speed, whereas \citet{huang_12} used a size-dependent relative velocity of the merging plasmoids, as dictated by resistive magnetohydronamic simulations of non-relativistic reconnection \citep[see also][]{huang_13}.
\begin{figure}
 \centering 
 \includegraphics[width=0.47\textwidth]{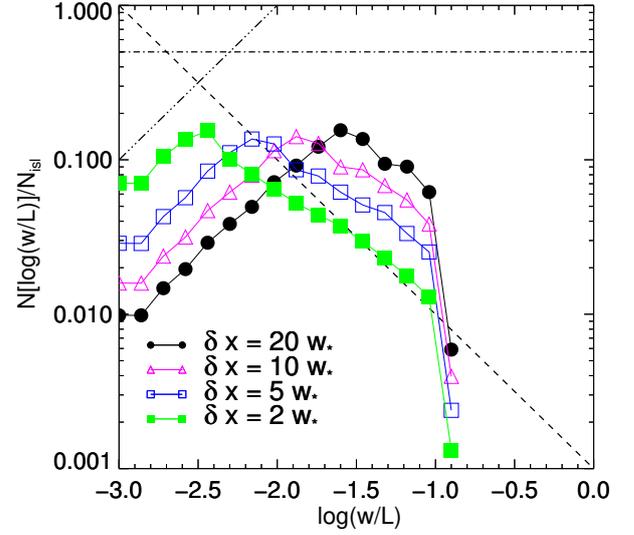}
\caption{Histogram of $\log (w/L)$ normalized to the total number of plasmoids, $N_{\rm isl}$, for different choices of the initial plasmoid separation distance $\delta x$, as indicated on the plot. All plasmoids move with the Alfv{\'e}n speed for $\sigma=10$, undergo mergers, and grow at a constant rate $\bg=0.4$. The dashed, dash-dotted, and dash triple-dotted lines correspond to the scalings $N(w)\propto w^{-2}$,  $N(w) \propto w^{-1}$, and $N(w)\propto$~const respectively.}
\label{fig:uzdensky}
\end{figure}

\begin{figure}
 \centering 
 \includegraphics[width=0.47\textwidth]{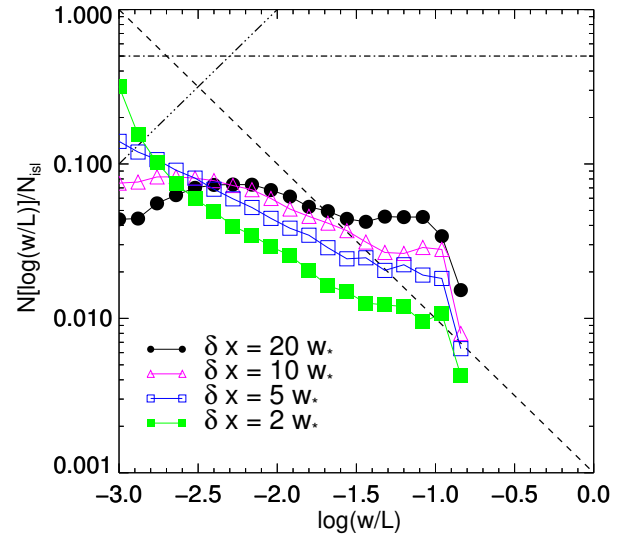}
\caption{Normalized histogram of $\log (w/L)$ for different choices of the initial plasmoid separation distance $\delta x$. Plasmoids move with constant speed drawn from a Gaussian distribution with zero mean and standard deviation equal to $v_{\rm A}/2$. Plasmoids undergo mergers, and grow at a constant rate $\bg=0.08$. Black lines have the same meaning as in Fig.~\ref{fig:uzdensky}.}
\label{fig:huang}
\end{figure} 

In order to compare our results with those presented by \citet{uzdensky_10} we (i) fix the plasmoid velocity in our MC code to the Alfv{\'e}n speed for $\sigma=10$, (ii) assume a constant growth rate in the co-moving frame of the plasmoids, (iii) set the separation distance of newly born plasmoids to be a multiple of the initial plasmoid size $w_*$, and let the system evolve for two light crossing times. By choosing  $\bg=0.4$ we ensure that plasmoids may grow at most up to $\sim \bg L /\sqrt{\sigma} \sim 0.1\,L$ for $\sigma=10$\footnote{For the same reason, we adjusted the value of $\bg$ in other numerical experiments, too.}. This results in a dynamic range of sizes  that is sufficient to explore the formation of a power law, as demonstrated by \citet{uzdensky_10}.

Our results are presented in Fig.~\ref{fig:uzdensky} for three choices of the separation distance $\delta x$ as indicated on the plot. The dashed line has slope $-1$ and corresponds to $N(w)\propto w^{-2}$. 
For each choice of $\delta x$ the size distribution exhibits two power law segments (i.e., $N(w)\propto$ const at small sizes and $N(w) \propto w^{-\chi}$ with $\chi \sim 1.7-1.9$ at intermediate sizes) followed by a steep cutoff. Our findings are in agreement with those presented by \citet{uzdensky_10}. Because all plasmoids in this scenario move with the same speed along the layer, mergers between neighboring plasmoids occur only if the plasmoids grow to a size comparable to their separation distance. As a consequence, the position of the break in the size distribution appears at $w \sim \delta x$, as illustrated in Fig.~\ref{fig:uzdensky}. Growth and advection from the layer are the two main processes that determine the evolution of plasmoids with $w \ll \delta x$ and, in this regime, $N(w)\propto$~const as mentioned by \citet{uzdensky_10} and \citet{huang_12}. At sizes $w\gg \delta x$, the competition of plasmoid growth and mergers leads to a steeper power law whose index is in rough agreement with the findings by \citet{uzdensky_10}. 
 
We next illustrate the effect of the plasmoid velocity distribution on the size distribution and comment on the scaling $N(w)\propto w^{-1}$ reported by \citet{huang_12}. To do so, we assume that plasmoids are born with velocity drawn from a Gaussian distribution with zero mean and standard deviation equal to $v_{\rm A}/2$. As long as the adopted distribution has an extent of the order of the Alfv{\'e}n speed, the exact functional form  does not affect the size distribution.  Furthermore, the plasmoid velocity remains constant in time and does not depend on the birth location in the layer. Our numerical setup is similar but not identical to that used in \citet{huang_12}. In their kinetic equation approach, the Gaussian profile referred to the relative velocity of merging plasmoids. This cannot be set as an independent parameter in our MC code, since it is automatically determined by the histories of the merging plasmoids. Additionally, the advection velocity of plasmoids from the layer was set equal to $v_{\rm A}$ by \citep{huang_12}. In our MC description of the plasmoid chain, the advection velocity is a parameter that is being determined by the histories of individual plasmoids. 

Figure~\ref{fig:huang} shows the histogram of $\log w$ as obtained from our MC code for different choices of the initial separation distance and $\bg=0.08$. The size distributions are qualitatively different from those shown in Fig.~\ref{fig:uzdensky}, thus highlighting the role of the plasmoids' relative motion in shaping the size distributions. The relative motion between plasmoids leads to more frequent mergers than in the scenario where all plasmoids move with the same speed. For $\delta x \gtrsim 10 w_*$, we find that $N(w) \propto w^{-\chi}$ with $\chi \sim 1.1-1.3$ (see black and magenta symbols). Given the fact that we cannot make the exact same assumptions as \citet{huang_12}, our result is in rough agreement with the scaling $\propto w^{-1}$ reported therein. However, the distribution tends to become softer for a given growth rate,  if the initial separation distance gets smaller (blue and green symbols in Fig.~\ref{fig:huang}). The slope of the differential size distribution lies between $-2$ and $-1$, i.e. between the values reported by \citet{uzdensky_10} and \cite{huang_12}, respectively. In this regime, plasmoids merge not only due to their relative motion but also due to their growth. Even if two neighboring plasmoids move with the same speed, they will eventually overlap because of their growth.  

Interestingly, our results support a scenario where $N(w) \propto w^{-2}$ for $\delta x \approx w_*$ and $N(w)\propto w^{-1}$ for $\delta x \gg w_*$, within the common simplifying assumptions made  by \citet{uzdensky_10} and \cite{huang_12} (i.e., constant plasmoid speed and constant growth rate).

\subsection{Analysis of individual processes}\label{sec:breakdown}
The focus of this section is the plasmoid acceleration and merging as well as their role in shaping the plasmoid size distribution. Armed with the MC code described in the previous sections, we may study in more detail individual processes. Readers who are not interested in the interplay of the various physical processes can skip this section and move directly to Sect.~\ref{sec:astro}.

\subsection{Effects of plasmoid acceleration}\label{sec:acceleration}
We first neglect mergers as a loss process for plasmoids in the layer. Yet, their effect is still included in the prescription for the plasmoid growth. Although switching off mergers in the MC code is an artificial choice (the plasmoids will effectively pass through each other), it helps us to highlight the role of the plasmoid bulk acceleration, a process that has been neglected in the models of \citet{uzdensky_10} and \citet{huang_12}. 

\begin{figure}
 \centering
 \includegraphics[width=0.47\textwidth]{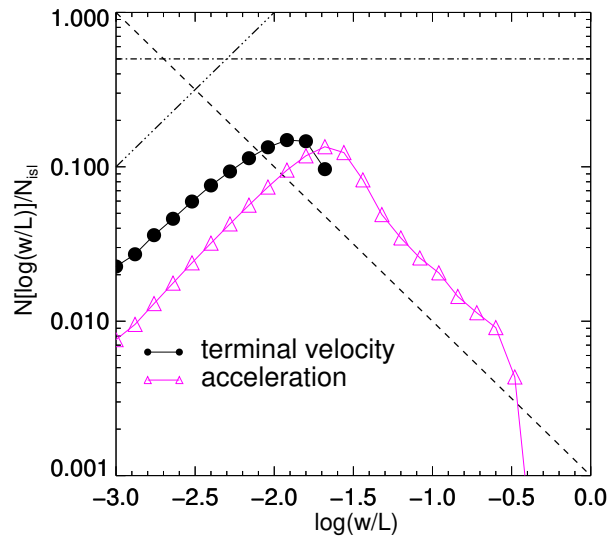}
 \caption{Normalized histogram of $\log w$ obtained from our MC code for $\sigma=10$ in two cases: all plasmoids move with the terminal velocity $v_{\rm A}$ (black symbols) or plasmoids accelerate according to eqn.~(\ref{eq:p}) (magenta symbols). Here, $\bg=0.08$ and $\bacc=0.12$ (Table~\ref{tab:param}).  Mergers, as a loss process for plasmoids, are not included here. The system is evolved for $6 L/c$. Black lines have the same meaning as in Fig.~\ref{fig:uzdensky}.}
 \label{fig:accel-1}
\end{figure}

Figure~\ref{fig:accel-1} shows the size distribution for $\sigma=10$ in two cases: plasmoids move with $v_{\rm A}$ (black symbols) and plasmoids accelerate according to eqn.~(\ref{eq:p}) (magenta symbols). The adopted growth and acceleration rates are presented in Table~\ref{tab:param}. There is also no suppression of growth for fast moving plasmoids (i.e., $f_{\rm sup}=1$). In the absence of mergers, plasmoids are only lost due to advection from the layer. The size distribution is $N(w) \propto$ const, when all plasmoids are moving with the asymptotic speed, in agreement with analytical arguments (see Appendix~\ref{app:app1} and \cite{uzdensky_10}). The distribution cuts off at a maximum size of $\wmx \approx \bg L /\sqrt{\sigma} \simeq 0.03\, L$ in agreement with analytical estimates (see eqn.~(\ref{eqn:wmax})). 

\begin{figure}
 \centering
 \includegraphics[width=0.47\textwidth]{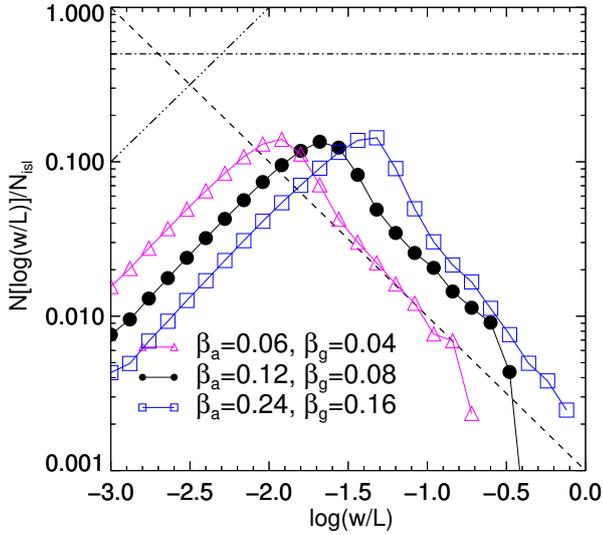}
 \caption{Normalized histogram of $\log w$ obtained with our MC code for $\sigma=10$.  Plasmoids accelerate according to eqn.~(\ref{eq:p}). Mergers as a loss process of plasmoids are neglected. The results are obtained for different choices of $\bacc$ and $\bg$ that retain the same ratio $\bacc/\bg$. We let the system evolve for $6L/c$. Black lines have the same meaning as in Fig.~\ref{fig:uzdensky}.}
 \label{fig:accel-2}
\end{figure} 
When the plasmoid acceleration is taken into account, a softer power law segment emerges above the characteristic size $\wmx$ (magenta symbols in Fig.~\ref{fig:accel-1}).  The formation of the second power law is attributed to the different four-velocities of individual plasmoids, which affect both their growth rates (see eqs.~(\ref{eq:growth}) and (\ref{eq:p})) and their residency time in the layer.  The size distribution is similar to the one derived by \citet{uzdensky_10},  albeit for different reasons. Here, it is the plasmoid acceleration and growth that produce the second power-law segment and not the plasmoid mergers. 

Using analytical arguments, one can show that a power law ($N(w) \propto w^{-\bacc/\bg}$) is expected for plasmoids with $w>\bacc L /\sqrt{\sigma}\gtrsim \wmx$ (see Appendix ~\ref{app:app1}). These plasmoids are advected from the layer before reaching their terminal velocity. 
The  size distribution obtained from the MC code (magenta symbols) is softer by 0.5 than the analytical prediction. This discrepancy is mainly caused by one of the simplifying assumptions made in our analytical approach, namely the omission of the term $p_0$ in eqn.~(\ref{eqn:f_approx}). Indeed, if we modify the MC code so that $|p_0| \ll \sqrt{\sigma}$ everywhere in the layer, we recover the power law index of $-\bacc/\bg$ predicted by our analytical study. 

Despite this discrepancy, we were able to verify that the slope of the second power law segment depends linearly on the ratio $\bacc/\bg$, as predicted analytically. Figure~\ref{fig:accel-2} shows the size distribution obtained  for different choices of $\bacc$ and $\bg$ which, however, retain the same ratio $\bacc/\bg$.  Indeed, the slope of the second power law segment $\chi$ is the same for all three cases (here, $\chi\simeq 2$).  In addition, we find that the break in the size distribution is  $\propto \bg$. This supports our previous interpretation that the break occurs at $\approx \bg L / \sqrt{\sigma}$ (see also Fig.~\ref{fig:accel-1}). Overall, the second power law segment appears shifted towards larger sizes as long as the growth and acceleration rates increase by the same amount.  

We reach similar conclusions for the other plasma magnetizations. SGP16 pointed out that the size distributions obtained from their PIC simulations for different magnetizations are similar (see also Fig.~\ref{fig:comparison-1}). {\sl A posteriori} this is not unexpected,  since the ratio $\bacc/\bg$ does not vary much for the values of $\sigma$ explored therein (see Table~\ref{tab:param}).  

 \begin{figure}
 \centering
 \includegraphics[width=0.47\textwidth]{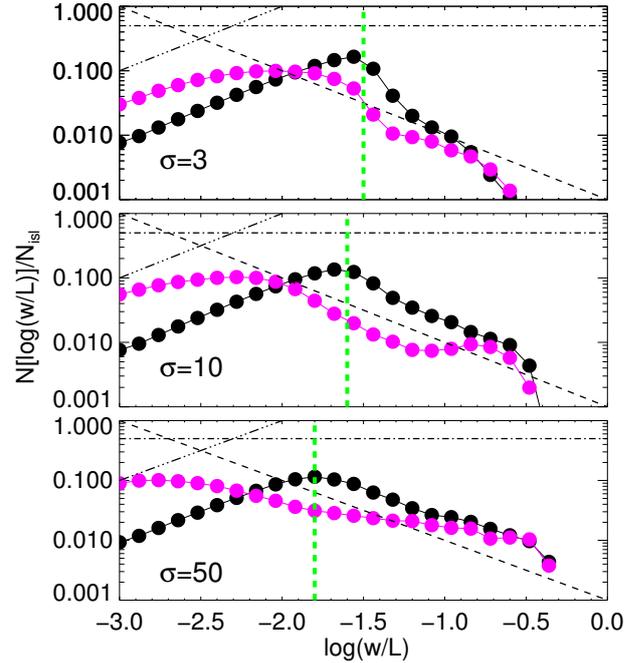}
 \caption{Normalized histogram of $\log w$ obtained with our MC code for $\sigma=3, 10$, and $50$ (top to bottom panels) when plasmoids accelerate and are lost due to  solely advection (black symbols) or due to both advection and merging (magenta symbols). The system is evolved for $6L/c$. The dashed green vertical line indicates  the plasmoid's characteristic size $\wmx=\bg L /\sqrt{\sigma}$. The values of $\bacc$ and $\bg$ are presented in Table~\ref{tab:param}. Black lines have the same meaning as in Fig.~\ref{fig:uzdensky}.  }
 \label{fig:merging-3}
\end{figure}  

\subsection{The role of mergers as a loss process}\label{sec:mergers}
We continue by including mergers as a loss process for plasmoids that accelerate in the layer according to eqn.~(\ref{eq:p}). At this point,  all of the relevant physical ingredients apart from the suppression of growth for fast plasmoids are taken into account. 

Although mergers do not  affect the momentum distribution (not shown), they have an impact on the size distribution, as illustrated in Fig.~\ref{fig:merging-3}   for $\sigma=3,10$, and 50 (top to bottom). The dashed vertical line indicates the characteristic size $\wmx$ above which a power law is formed due to the interplay of plasmoid acceleration and growth, as discussed in Sect.~\ref{sec:acceleration}.

For all plasma magnetizations, we find that the distribution at the largest sizes (i.e., $w\gtrsim 0.1\, L$) is not altered when mergers are taken into account, since the largest plasmoids are those that survive the mergers with smaller plasmoids.   On the contrary, faster plasmoids of intermediate sizes ($w \sim 0.01 -0.1 \, L$) are removed from the layer due to mergers with slower bigger plasmoids that lie ahead of them. There is also an excess of small plasmoids with $w \lesssim 0.01 \, L$ when mergers are included. This may seem counterintuitive at first sight. Yet, mergers create, at every time step,  free space between neighboring plasmoids which can be filled by new plasmoids with small sizes $\sim w_*$.

In the absence of mergers (black symbols) the size distribution, $N(\log w) \propto w N(w)$, can be approximated by a broken power law with a  break at $\approx \wmx$ (dashed vertical green line) that scales with the system size $L$. Mergers do not affect the broken power-law shape of the distribution, but they push the position of the break to smaller sizes (magenta symbols).  As we show in Sect.~\ref{sec:size}, the position of the break (henceforth noted as $w_{\rm m}$) is a  multiple of $w_*$ and, in contrast to $\wmx$, does not depend on the system's size. Fig.~\ref{fig:merging-3} also demonstrates that $w_{\rm m} \rightarrow w_*$, as the plasma magnetization increases. In fact, the break appears at $\sim w_*$ when $\sigma=50$, thus making the distribution look like a single power law (see also Fig.~\ref{fig:comparison-1}). 

\begin{figure}
 \centering
 \includegraphics[width=0.47\textwidth]{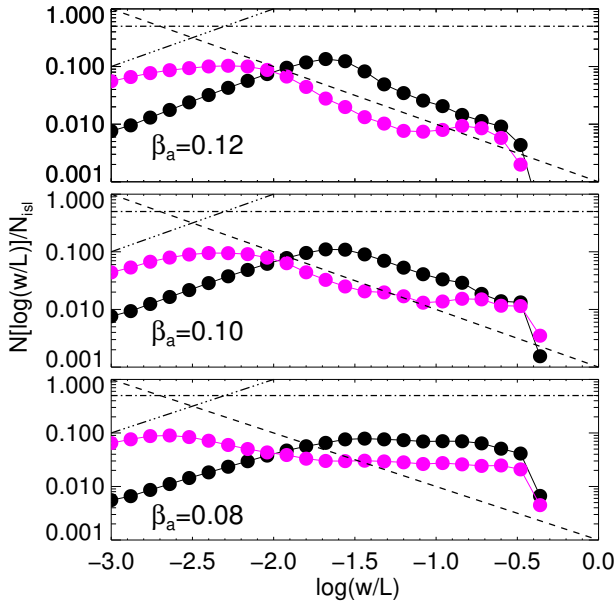}
  \caption{Same as Fig.~\ref{fig:merging-3} but for three fiducial values of the acceleration rate $\bacc$ that are indicated on the plot. } 
 \label{fig:merging-slope}
\end{figure}

The results presented in Fig.~\ref{fig:merging-3} suggest that mergers do not affect the slope of the second power law segment  (i.e., at $w\gtrsim w_{\rm m}$). We demonstrated that the latter depends on the ratio $\bacc/\bg$ (Sect.~\ref{sec:acceleration}). To investigate if the power law slope of the distribution is affected by mergers, we varied the acceleration rate and computed the corresponding size distributions. Our results are presented in Fig.~\ref{fig:merging-slope}. It is evident that the slope of the second power law segment (for plasmoids smaller than the monster ones) changes with $\bacc$, as discussed in Sect.~\ref{sec:acceleration}, and that it is not affected by the coalescence of plasmoids.  

\begin{figure}
 \centering
 \includegraphics[width=0.47\textwidth]{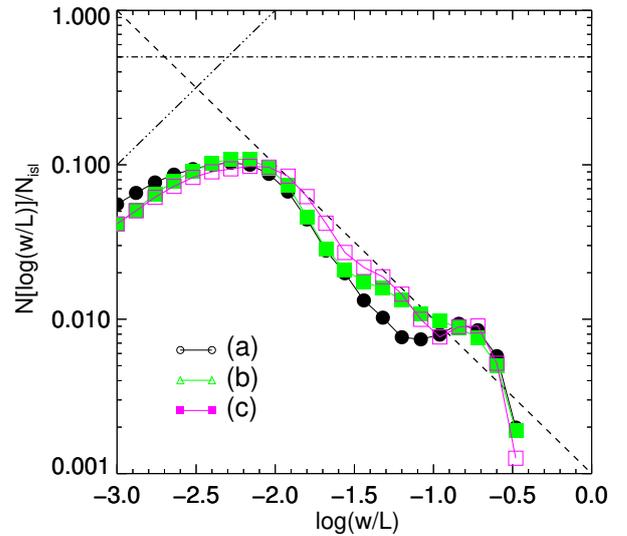}
 \caption{Normalized histogram of $\log w$ obtained with our MC code for $\sigma=10$ and three different prescriptions for plasmoid merging, as defined  in Sect.~\ref{sec:method}:  (a) intersection of outer boundaries, (b) intersection of centers, and (c) intersection of inner or outer boundaries.  We let the system evolve for $6L/c$. Black lines have the same meaning as in Fig.~\ref{fig:uzdensky}.}
 \label{fig:merging-4}
\end{figure}

So far, the identification of a merger in the MC code was based on the relative location of the plasmoids outer boundaries (for different merging criteria, see Sect.~\ref{sec:method}). Our results, however, are not sensitive to this choice, as demonstrated in Fig.~\ref{fig:merging-4}, where the size distribution is shown for three different prescriptions for the plasmoid mergers (for details, see Sect.~\ref{sec:method}).  

\begin{figure}
 \centering 
\includegraphics[width=0.47\textwidth]{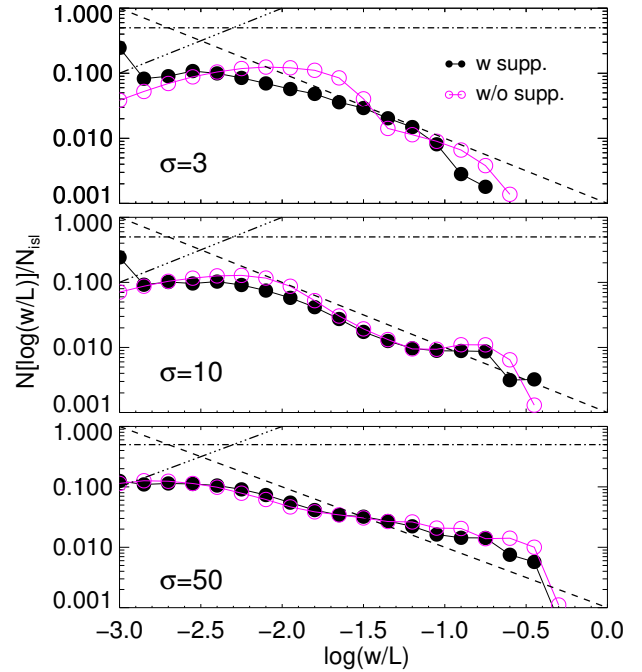} 
\caption{Normalized histogram of $\log w$ obtained with our MC code for a duration of $6L/c$ with (black filled symbols) and without (magenta open symbols) growth suppression, for $\sigma=3$ to 50 (from top to bottom). The magenta coloured histograms are the same as those displayed in Fig.~\ref{fig:merging-3}.   For both magenta and black symbols, plasmoids accelerate according to eqn.~(\ref{eq:p}), merge, or leave the layer. For the adopted values of $\bg$ and $\bacc$, see Table~\ref{tab:param}. Black lines have the same meaning as in Fig.~\ref{fig:uzdensky}.}
\label{fig:size_supp}
\end{figure} 

\subsection{The role of growth suppression}\label{sec:suppression}
The results of the SGP16 PIC simulations suggest that the growth rate of plasmoids moving with velocities that approach the Alfv{\'e}n speed is suppressed compared to the average growth rate of slower plasmoids in the chain (see also Appendix~\ref{app:app0}). The suppression of the growth is typically relevant for the smallest plasmoids\footnote{At small sizes (i.e., below $\wmx$) the most relevant processes for the plasmoids are growth and advection from the layer, as discussed in Sect.~\ref{sec:comparison}.}, as these have $|p|\approx \sqrt{\sigma}$. 

Fig.~\ref{fig:size_supp} shows the size distribution obtained with (black filled symbols) and without (magenta open symbols) growth suppression for $\sigma=3$ to 50 (from top to bottom). Growth suppression does not seem to affect the size distribution for $\sigma=50$ (bottom panel), since the majority of the plasmoids moves with $v < v_{\rm A}$, as shown in Fig.~\ref{fig:comparison-2} and Fig.~8 in SGP16. In particular, it remains true that most of the plasmoids for $\sigma=50$ lie at the smallest sizes  and the distribution resembles a single power law. For $\sigma=3$ and 10, we find that growth suppression affects the distribution of the smallest plasmoids (i.e., $w \lesssim w_{\rm m}$) since these are, in general, the fastest plasmoids in the layer. Growth suppression affects the size distribution in two ways. Firstly, the majority of the plasmoids has $w\approx w_*$, since all newly formed plasmoids with $w\sim w_*$ and $v \approx v_{\rm A}$ cannot grow much. Secondly, the characteristic break size of the distribution shifts towards smaller sizes, i.e. $w_{\rm s} \simeq 4  w_* \lesssim w_{\rm m}$ (see also the following subsection).

In summary, when growth suppression is taken into account, the size distribution at $w\gtrsim 4 w_\star$ is a single power law for all the magnetizations we explore, with power-law index that decreases from $\sim 2$ to $\sim 1.3$ as $\sigma$ increases from 3 to 50 (see also Sect.~\ref{sec:acceleration}).

We showed that the suppression can be phenomenologically described by eqn.~(\ref{eq:sup}), with the parameter $A$ being essentially a threshold of the plasmoid's four-velocity. Its dependence on $\sigma$ is weak, as shown in Table~\ref{tab:param}, and can be modelled as $A(\sigma) = \sigma^{-z}$ where $z\simeq 0.11$. The plasmoid four-velocity at this threshold is then $p_{\rm th}\equiv A\sqrt{\sigma}$. This corresponds to a dimensionless plasmoid velocity  of $v_{\rm th}\equiv p_{\rm th} c /\sqrt{p_{\rm th}^2+1}$ that is, in good approximation, a constant fraction of the Alfv{\'e}n speed (i.e., $\sim 0.96-0.98 \, v_{\rm A}$) for $\sigma=3-50$. This suggests that the accretion of trailing fast plasmoids, which is responsible for most of the plasmoid's growth at low  speeds,  is quenched when the growing plasmoid is also fast with $v\rightarrow v_{\rm A}$. 

\subsection{Effects of the system's size}\label{sec:size}
The length of the reconnection layer can widely vary depending upon the astrophysical environment. SGP16 demonstrated with large scale PIC simulations of relativistic reconnection that the plasmoid growth rate $\bg$ and acceleration rate $\bacc$ do not depend on the system's length. Similarly, the fluid properties of the plasmoids, such as plasma density and magnetic energy density,  were found to be the same in systems with different sizes.  In this section, we discuss the effects of the system's size on the statistical properties of the plasmoid chain.

\begin{figure}
 \centering 
\includegraphics[width=0.47\textwidth, trim=20 100 0 0]{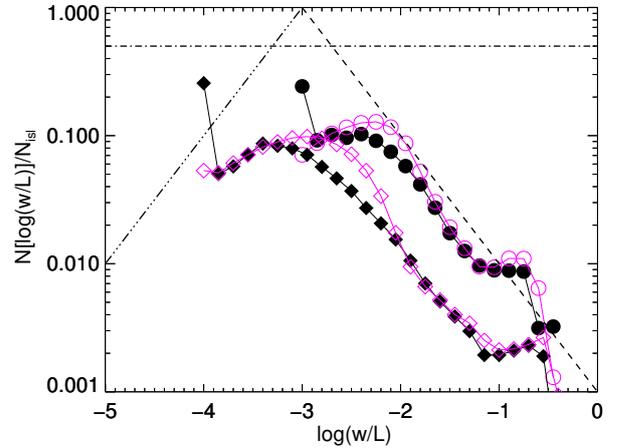} 
\caption{Normalized histogram of $\log w$ obtained with our MC code for $\sigma=10$ and two values of the ratio $w_*/L$, namely $10^{-3}$ (circles) and $10^{-4}$ (diamonds). Black filled symbols show the distribution when all processes are included. Histograms plotted with magenta open symbols show the distribution when  growth suppression is neglected. Histograms plotted with circles are the same as in Fig.~\ref{fig:size_supp}. The rates $\bg$ and $\bacc$ are fixed to their nominal values (Table~\ref{tab:param}) and black lines have the same meaning as in Fig.~\ref{fig:uzdensky}.}
\label{fig:extend}
\end{figure}

Let us consider systems with different sizes and let $L$ denote the half-length of the layer. Then, the ratio $w_*/L$ is a measure of the system's size, since the plasmoid size at birth is, in all cases, equal to a few electron skin depths. Fig.~\ref{fig:extend} shows the size  distribution of plasmoids for $\sigma=10$ and systems with $w_*/L=10^{-3}$ (circles) or $10^{-4}$ (diamonds). 

In both cases, growth suppression affects the size distribution the same way (for details, see Sect.~\ref{sec:suppression}). Our results demonstrate the formation of a power law  whose slope does not depend on the growth suppression. Its dynamic range becomes wider for larger systems. On the one hand, the  power law cuts off at a maximum plasmoid size that is a fraction (10-30 per cent) of the layer's length. On the other hand, the power law develops above a characteristic size $w_{\rm br}$ that is a constant multiple of the plasmoid's size at birth. The break size $w_{\rm br}$ can identified as $w_{\rm m}$ in the absence of growth suppression (see Sect.~\ref{sec:mergers}) or  $w_{\rm s} \lesssim w_{\rm m}$ when growth suppression is taken into account (see Sect.~\ref{sec:suppression}).  We also find no difference between the momentum distributions (not shown), since the prescriptions for the plasmoid acceleration (eqn.~(\ref{eq:p})) and initial momentum (eqn.~(\ref{eq:p0})) do not depend on the system's size. Similar calculations can be performed for astrophysical systems where the separation between the plasma scales and the system's size is $\gtrsim 10^5$ (e.g., flaring PWNe).

\subsection{Distributions of exiting plasmoids}\label{sec:exit}
So far, our analysis focused on the time- and position-integrated statistical properties of the plasmoid chain. Yet, only a fraction of the plasmoids that form in the layer at any time will be able to exit before merging. How does the distribution of plasmoid sizes and momenta at the time of their exit from the layer compare to the integrated distributions? In a steady state system the two distributions should be identical provided that the properties of the plasmoid chain are uniform across the layer. However, the PIC results recently presented by SGP16 suggest that the properties of individual plasmoids may depend on their location in the layer (see also Sect.~\ref{sec:pic}). 

\begin{figure}
 \centering 
\includegraphics[width=0.47\textwidth]{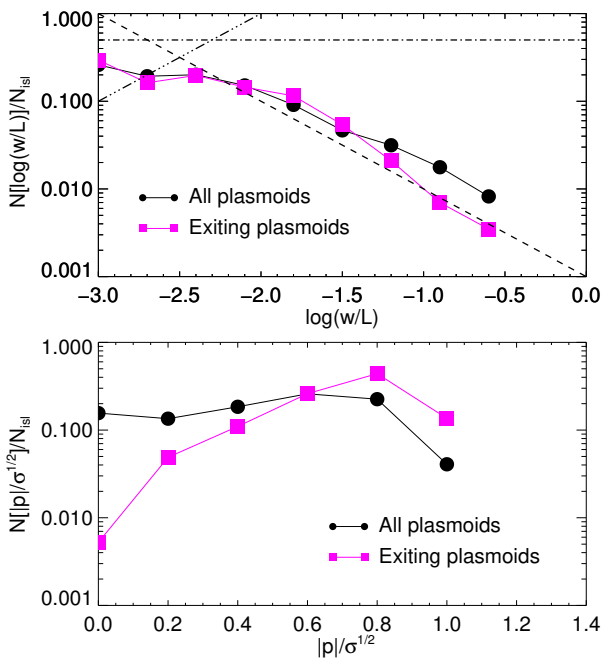} 
\caption{Size (top panel) and momentum (bottom panel) distributions of all plasmoids that have been present  in the layer (black symbols) and of plasmoids that leave the layer (magenta symbols). The results are obtained for $\sigma=10$ and $w_*/L=10^{-3}$. All processes, including the growth suppression, are taken into account. The duration of the MC simulations was fixed to $6 L/c$.}
\label{fig:exit}
\end{figure} 

Figure~\ref{fig:exit} shows the integrated size and momenta distributions (black symbols in top and bottom panels, respectively) of the plasmoid chain from a MC simulation with $\sigma=10$. The distributions of plasmoid sizes and momenta at the time of their exit from the layer is also shown for comparison (magenta symbols). We find that the two size distributions match for small and intermediate plasmoid sizes (i.e., $w \lesssim 0.03 \, L$). The integrated size distribution shows an excess for $w \gtrsim 0.03 \, L$, which is caused by the presence of slow moving plasmoids in the layer. These have not yet reached the edge of the layer and are, thus, not counted in the magenta coloured histogram. The exiting momentum distribution is qualitatively different from the integrated one, as presented in the bottom panel of Fig.~\ref{fig:exit}.  Plasmoids that have low four-velocities are typically born close to the center of the layer (see Fig.~\ref{fig:init-dist}).  
Those that will eventually exit the layer  will have been accelerated to higher four-velocities by that time. Additionally, plasmoids are born fast, if their birth location is closer to the edge of the layer (see Fig.~\ref{fig:init-dist}).

\section{Blazar variability}\label{sec:astro}
Although blazars constitute a  small subclass of AGN, they are discovered in increasingly large numbers by surveys at microwave wavelengths and $\gamma$-ray energies \citep[e.g.][]{giommiWMAP_09, abdo_10, giommiPlanck12, ackermann_15}. Blazars also represent  the most abundant population of extragalactic sources at TeV energies\footnote{http://tevcat.uchicago.edu/} \citep[e.g.][]{holder_ICRC13, deNaurois_ICRC15}, suggesting particle acceleration up to very high energies.
The extreme observational properties of blazars, such as continuum emission over the entire electromagnetic spectrum, rapid and large-amplitude variability, make them to stand out among other AGN.  The blazar broadband emission, from radio up to very high energy $\gamma$-rays ($>$100~GeV),
is believed to originate from a relativistic jet that is nearly aligned with the observer's line of sight and emerges from the central supermassive black hole 
\citep{blandfordrees78, urry_padovani95}. 

Jets are likely to be launched as Poynting-flux dominated flows \citep[e.g.][]{blandford_znajek77, blandford_payne82} and remain such after their acceleration and collimation \citep[e.g.][]{spruit_96, vlahakis_04}. In regions where the magnetic field changes polarity, energy can be dissipated and eventually transferred to radiating particles via magnetic reconnection.

\cite{spg_15} have recently demonstrated that magnetic reconnection can   satisfy all the basic conditions for the blazar emission: efficient dissipation, extended (in energy) particle distributions, and rough energy equipartition between particles and magnetic fields in the plasmoids, which are identified as the emitting regions in blazar jets. Plasmoids may also naturally reproduce the extreme energetics and timescales of the observed flaring episodes
in blazars, as proposed by \cite{giannios_13}.  

Using the PIC results of SGP16, \cite{pgs_16} (henceforth, PGS16) derived analytical estimates for two fundamental observables of flares powered by plasmoids, namely the flux-doubling timescale $\Dtdb$ and the peak luminosity $L_{\rm pk}$. Both quantities are expressed in terms of the plasmoid's size ($w_{\rm f}$) and Doppler factor ($\delta_{\rm f}$) at the end of its lifetime\footnote{For the definition of the plasmoid's Doppler factor, see eqn.~(8) in PGS16.}, as shown below:
\eqb
\label{eq:dt12}
\Delta t_{1/2}  \approx \frac{w_{\rm f}}{\delta_{\rm f} \bg c} 
\eqe
and 
\eqb
\label{eq:Lbol}
L_{\rm pk} \approx \frac{f_{\rm rec} L_{\rm j}}{8\varpi^2 c \beta_{\rm j} \Gamma_{\rm j}^2}\bg c w_{\rm f}^2 \delta_{\rm f}^4 ,
\eqe
where $L_{\rm j}$ is the absolute jet power, $\Gamma_{\rm j}=\left(1-\beta_{\rm j}^2\right)^{-1/2}$ is the jet's bulk Lorentz factor, $\varpi$ is the jet's cross sectional radius, and $f_{\rm rec}$ is the fraction of energy that is transferred to radiating particles by reconnection. Before applying these  analytical results to the plasmoid chain derived from our MC simulations, we summarize a few caveats entering the calculation by PGS16:
\begin{itemize}
\item The suppression of the growth was not taken into account.
 \item The expression for the peak bolometric luminosity was derived under the assumption that the radiating particles are fast cooling, which is, in general, true for the largest plasmoids in the layer. For smaller plasmoids, expression (\ref{eq:Lbol}) gives an upper limit to the peak flare luminosity.
 \item Both expressions were derived assuming that the peak flux is always reached at the end of a plasmoid's lifetime, when its velocity is also maximum. This assumption is always valid  for non-relativistic plasmoids. For the smaller and faster (relativistic) plasmoids though, our assumption holds  as long as there is {\it optimal orientation} between the direction of the plasmoid's motion and the observer's line of sight.  For any other alignment, the emission produced by the smaller and faster plasmoids may peak before they reach their final size. 
\end{itemize}

\begin{figure}
 \centering
 \includegraphics[width=0.47\textwidth]{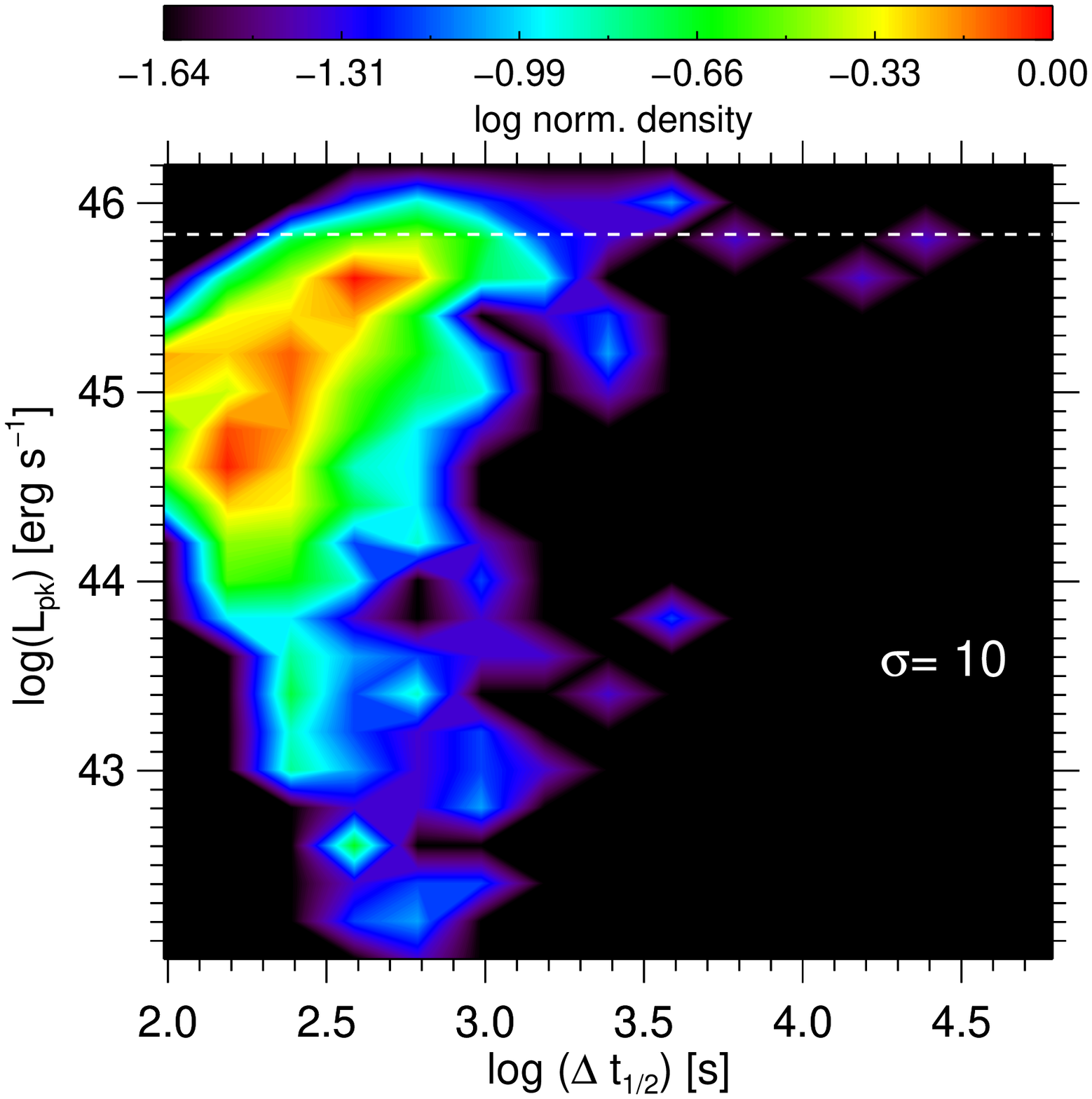}
  \includegraphics[width=0.47\textwidth]{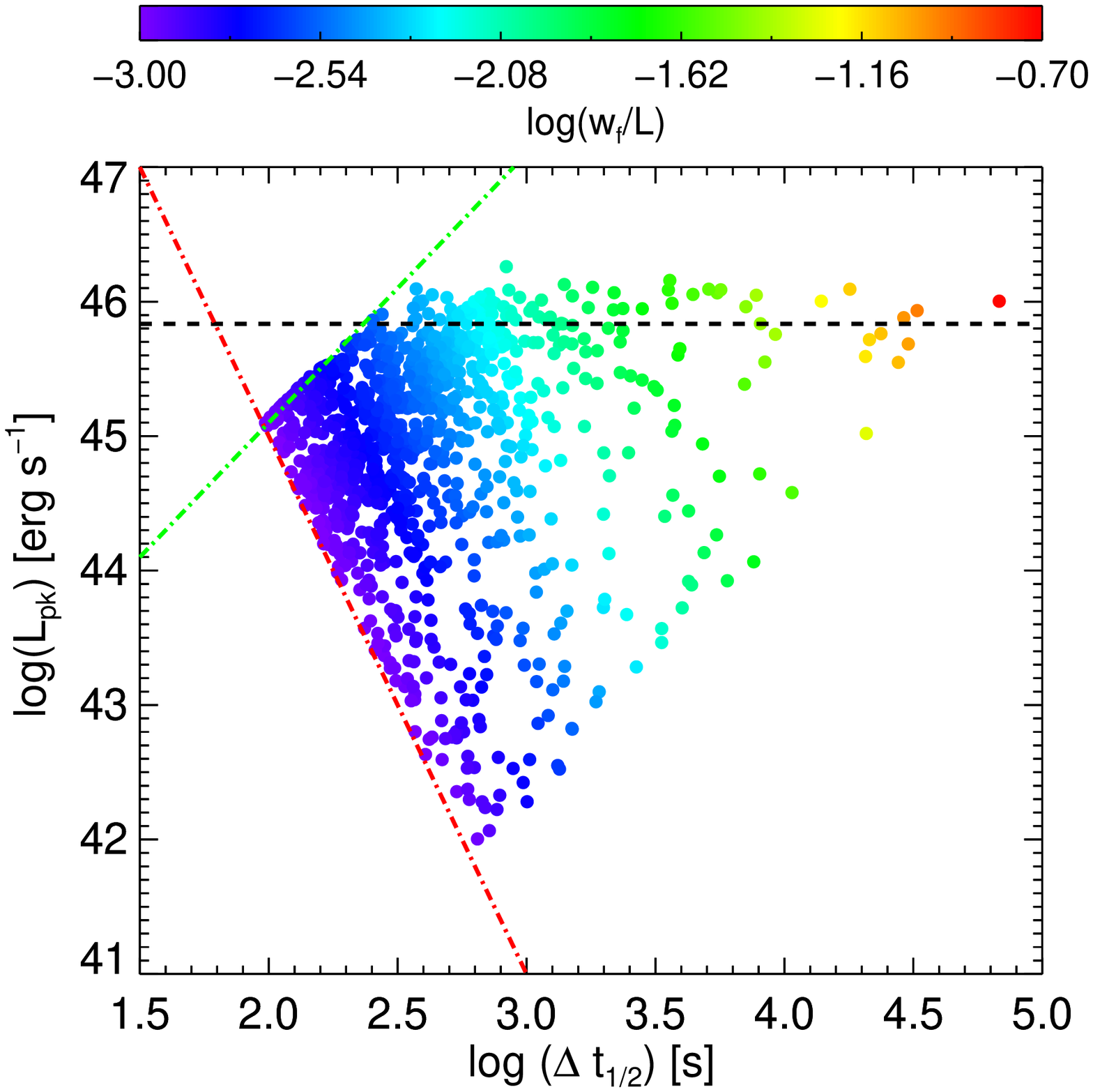}
 \caption{Density map (top panel) and scatter plot (bottom panel) of the peak bolometric luminosity and flux-doubling timescale of flares powered by the plasmoid chain from a MC simulation with $\sigma=10$ and $w_*/L=10^{-3}$. In the scatter plot a colour coding according to the final plasmoid size was used. The white and black dashed line in the top and bottom panels, respectively, indicates $L_{\rm av}/3$ (see eqn.~(\ref{eq:Lav})). The peak luminosity of flares powered by plasmoids of fixed final size scales as $\Delta t_{1/2}^{-4}$ (red dash-dotted line in bottom panel). Plasmoids with fixed final Doppler factor power flares with
 $L_{\rm pk} \propto \Delta t_{1/2}^2$ (green dash-dotted line in bottom panel).}
 \label{fig:lum_dt}
\end{figure}

Figure~\ref{fig:lum_dt} shows the density map (top panel) and the scatter plot (bottom panel) of the peak bolometric luminosity and flux-doubling timescale of flares powered by the plasmoid chain from a MC simulation with $\sigma=10$. Other parameters used are: $L_{\rm j}=10^{46}$~erg s$^{-1}$, $\Gamma_{\rm j}=10$, $f_{\rm rec}=0.5$, and $\varpi=10^{16}$~cm. The half-length of the layer is taken to be equal to $\varpi$. The average luminosity of  the reconnection event can be also estimated as:
\eqb 
L_{\rm av} \approx \frac{1}{T}\sum_i^{N_{\rm isl}} L_{\rm pk, i} \Delta t_{1/2, i}
\label{eq:Lav}
\eqe 
where $T$ is the observed duration of the event and is given by $L /\beta_{\rm rec} c\delta_{\rm j} \approx 2.5$~days, where $\delta_{\rm j}\approx 2\Gamma_{\rm j}$ is the jet's doppler factor and $\beta_{\rm rec}\simeq 0.08$ is the reconnection rate. Overplotted in the top and bottom panels of Fig.~\ref{fig:lum_dt} with white and black dashed lines, respectively, is $L_{\rm av}/3$. Using eqs.~(\ref{eq:dt12}) and (\ref{eq:Lbol}) one can also show that the peak luminosity of flares powered by plasmoids of fixed $w_{\rm f}$ depends on the flux-doubling timescale as $L_{\rm pk} \propto \Delta t_{1/2}^{-4}$ -- see red dash-dotted line in the bottom panel of Fig.~\ref{fig:lum_dt}. Plasmoids with fixed final Doppler factor but different $w_{\rm f}$ power flares with $L_{\rm pk} \propto \Delta t_{1/2}^2$ (green dash-dotted line in bottom panel in Fig.~\ref{fig:lum_dt}).

Most of the flares powered by the plasmoid chain have short durations and peak luminosities below $L_{\rm av}/3$ (see yellow and red coloured regions in the top panel of Fig.~\ref{fig:lum_dt}) and only a small fraction of flares ($\sim 8\%$) has $L_{\rm pk} > L_{\rm av}/3$. These will most likely be resolved as individual short-duration flares, whereas the superposition of many and less luminous flares will form an envelope of longer duration \citep[see also][]{giannios_13}. Detailed light curves from the plasmoid chain will be presented elsewhere (Christie et al., in prep.).

In this example, the single reconnection event lasts for several days and results in  many bright ($L_{\rm pk}>10^{46}$ erg s$^{-1}$) flares with durations ranging from $\sim$5~min to several hours. We find that long duration flares (here with $\Delta t_{1/2} \gtrsim 10^4$~s) are also luminous ($L_{\rm pk} \gtrsim L_{\rm av}/3$) and are produced by the largest plasmoids of the chain ($w_{\rm f} \gtrsim 0.03 \, L$), in agreement with PGS16. Plasmoids of intermediate sizes ($0.003 \, L \lesssim w_{\rm f} \lesssim 0.03 \, L$) produce shorter duration flares with a wide range of luminosities, while the smallest plasmoids ($w<0.003\, L$) power flares that are typically less luminous (i.e., $L_{\rm pk}<L_{\rm av}/3$).
 
\section{Summary \& Discussion}\label{sec:summary}
Magnetic reconnection is a highly dynamical process which leads to the formation of self-similar structures containing magnetic fields and energetic particles, the so-called plasmoids. These can merge with each other, grow in size, accelerate due to magnetic tension forces, and advect out of the layer. 

We  have developed a MC code to study the effects of these physical processes on the size and momentum distributions of plasmoids, in light of recent results from large-scale PIC simulations of relativistic magnetic reconnection. 

 We showed that the differential plasmoid size distribution forms a power law, $N(w)\propto w^{-\chi}$,  beyond a characteristic size $w_{\rm br}$ that is a constant multiple of the plasmoid's size at birth $w_*$. The break size shifts closer to $w_*$ for higher magnetizations, thus making the size distribution for $\sigma=50$ to appear like a single power law. In general, we find that the power law segment above $w_{\rm br}$ extends from small  sizes (i.e., few to tens of plasma skin depths) up to large sizes (i.e., a few percent of the reconnection layer's length). The slope of the power law $\chi$ above $w_{\rm br}$ decreases from $\sim 2$ to $\sim 1.3$ as  $\sigma$ increases from 3 to 50. We demonstrated numerically and analytically that the slope depends linearly on the ratio of the plasmoid acceleration and growth rates. The differential distribution of plasmoid momenta becomes softer at higher $\sigma$, in agreement with PIC results, and does not depend on the system's size. 

The MC code we presented in this paper was developed based on the findings of PIC simulations of relativistic magnetic reconnection. In principle, a similar analysis could be performed for the non-relativistic case with application to, e.g., accretion discs. The MC code is also flexible, for it can be easily modified to account for additional forces acting upon the plasmoids, such as the inverse Compton drag force induced by strong radiation fields. These could be either external to the reconnection layer or produced by the plasmoid chain itself \citep[see e.g.][]{belo_17}. 

The MC approach facilitates also the study of the plasmoid chain in large spatial domains. This is important for bridging the separation of the microscopic plasma scales with the macroscopic astrophysical scales and making meaningful predictions for the temporal properties of non-thermal radiation. 
Not requiring the use of computationally heavy PIC simulations, the results of the MC approach can be directly mapped to the statistical properties of flaring astrophysical sources, such as blazars and PWNe. 

\section*{Acknowledgements} 
We  acknowledge support from NASA through the grants NNX16AB32G and NNX17AG21G issued through the Astrophysics Theory Program.  LS acknowledges support from DoE DE-SC0016542, NASA Fermi NNX16AR75G, and NSF ACI-1657507.


\bibliographystyle{mnras} 
\bibliography{pic.bib} 

\appendix
\section[]{Growth suppression in PIC and MC simulations}
\label{app:app0} 
The requirement for a growth suppression factor, briefly discussed in Sect.~\ref{sec:pic}, arises from the overestimate of the plasmoids' sizes, if we were to adopt our first order approximation for the plasmoid growth (see eqn.~(\ref{eq:growth})).  This is exemplified in Fig.~8 in SGP16. The co-moving growth rate of the largest and longest-living plasmoids can be accurately considered to be constant. These appear as straight lines with a slope equal to $\bg$ in a plot of the plasmoid size versus co-moving time (Fig.~8 in SGP16). On the other hand, the small plasmoids, which are typically born with high momenta, do not follow this trend and quickly deviate away from a linear growth. 
\begin{figure}
 \centering
 \includegraphics[width=0.45\textwidth]{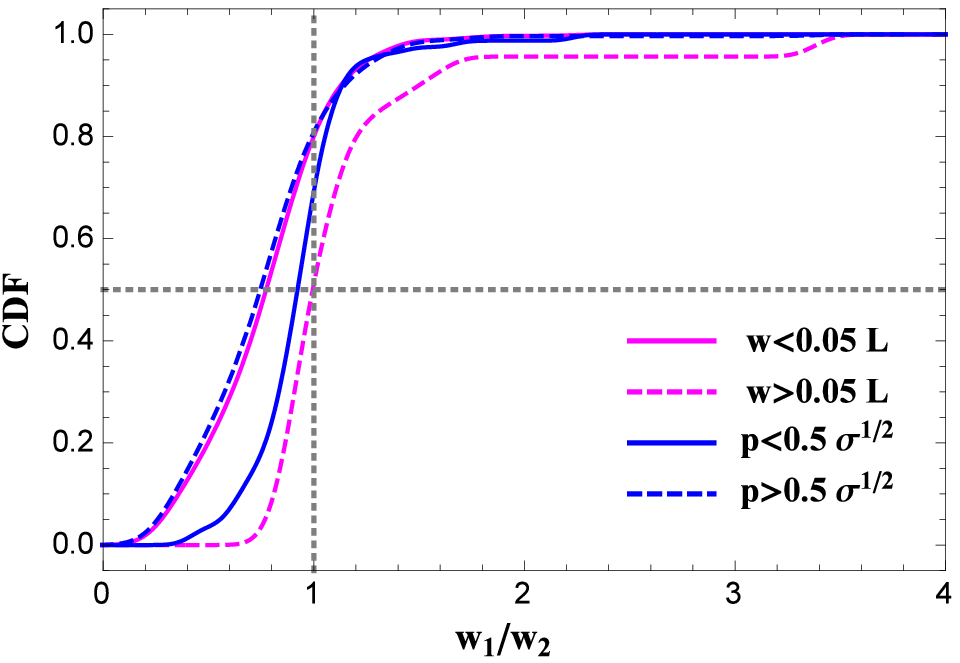}
 \caption{Cumulative distribution function (CDF) of the ratio of plasmoids final sizes as obtained from PIC data $w_1$ over that obtained using eqn.~(\ref{eq:growth}), $w_2$. The CDF was calculated for two sub-samples of the data. The sample was divided according to the plasmoids' final size (magenta coloured lines) or momentum (blue coloured lines).}
 \label{fig:suppression}
\end{figure}

This discrepancy  can also be seen in Fig.~\ref{fig:suppression}. Here, we plot the cumulative distribution function (CDF) of the ratio of a plasmoid’s final size as determined in PIC, $w_1$, to the final size predicted from eq.~(\ref{eq:growth}), $w_2$, for a constant growth rate $\beta_{\rm g}=0.08$. 
For those plasmoids who have either low final momenta or large final size, we find a symmetric distribution centered at unity, i.e., similar to the CDF of a Poisson distribution. However, for plasmoids who have high momenta or a small final size, we find the distribution is shifted below unity suggesting that eqn.~(\ref{eq:growth}) is over predicting the plasmoid’s growth. Our choice for the suppression factor is given by eqn.~(\ref{eq:sup}). In the following section, we discuss our method of best determining its free parameters as a function of plasma magnetization.
\subsection[]{Calibration of the suppression factor}\label{sec:calibration}
\begin{figure}
 \centering 
\includegraphics[width=0.4\textwidth]{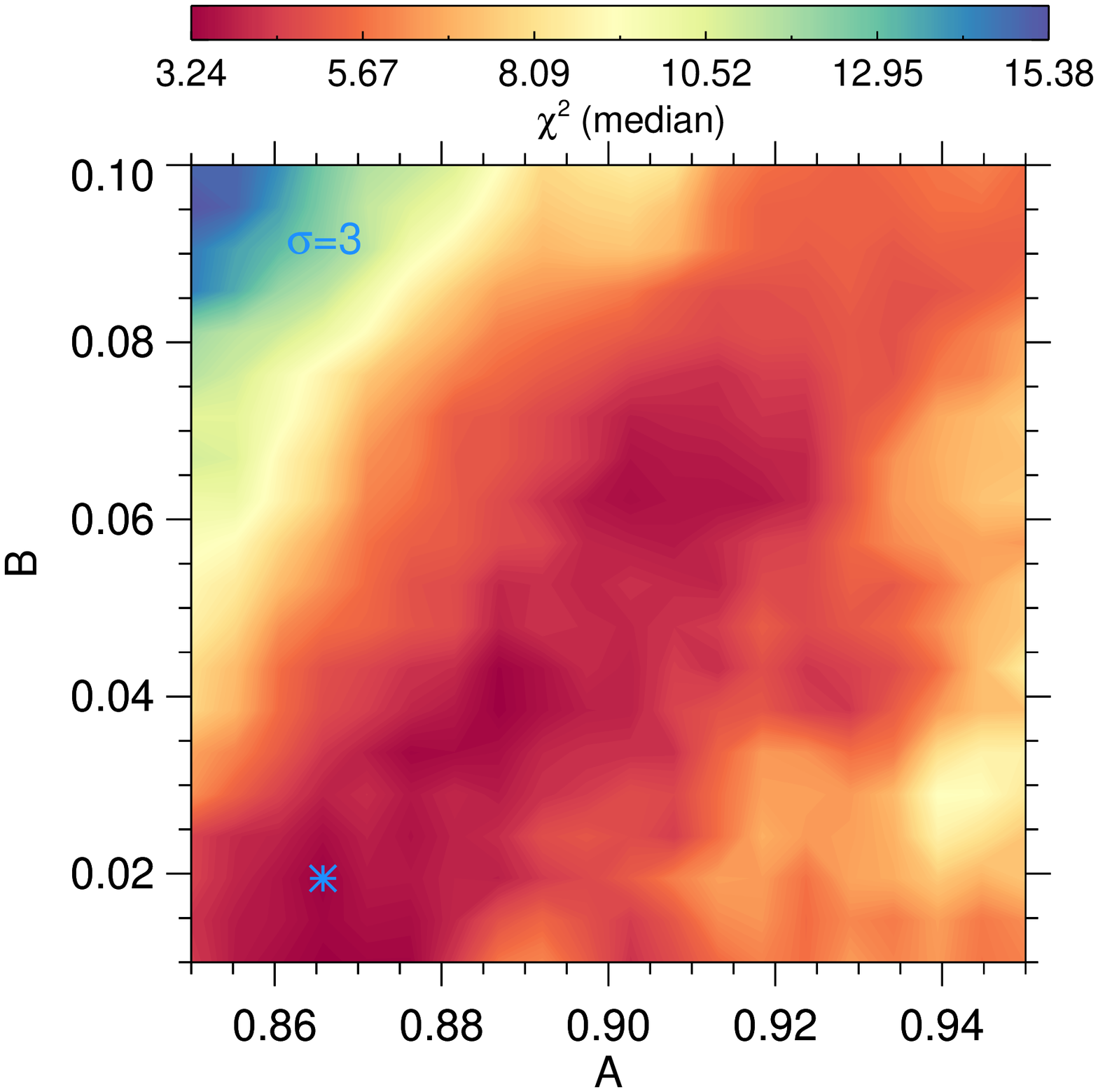} 
\includegraphics[width=0.4\textwidth]{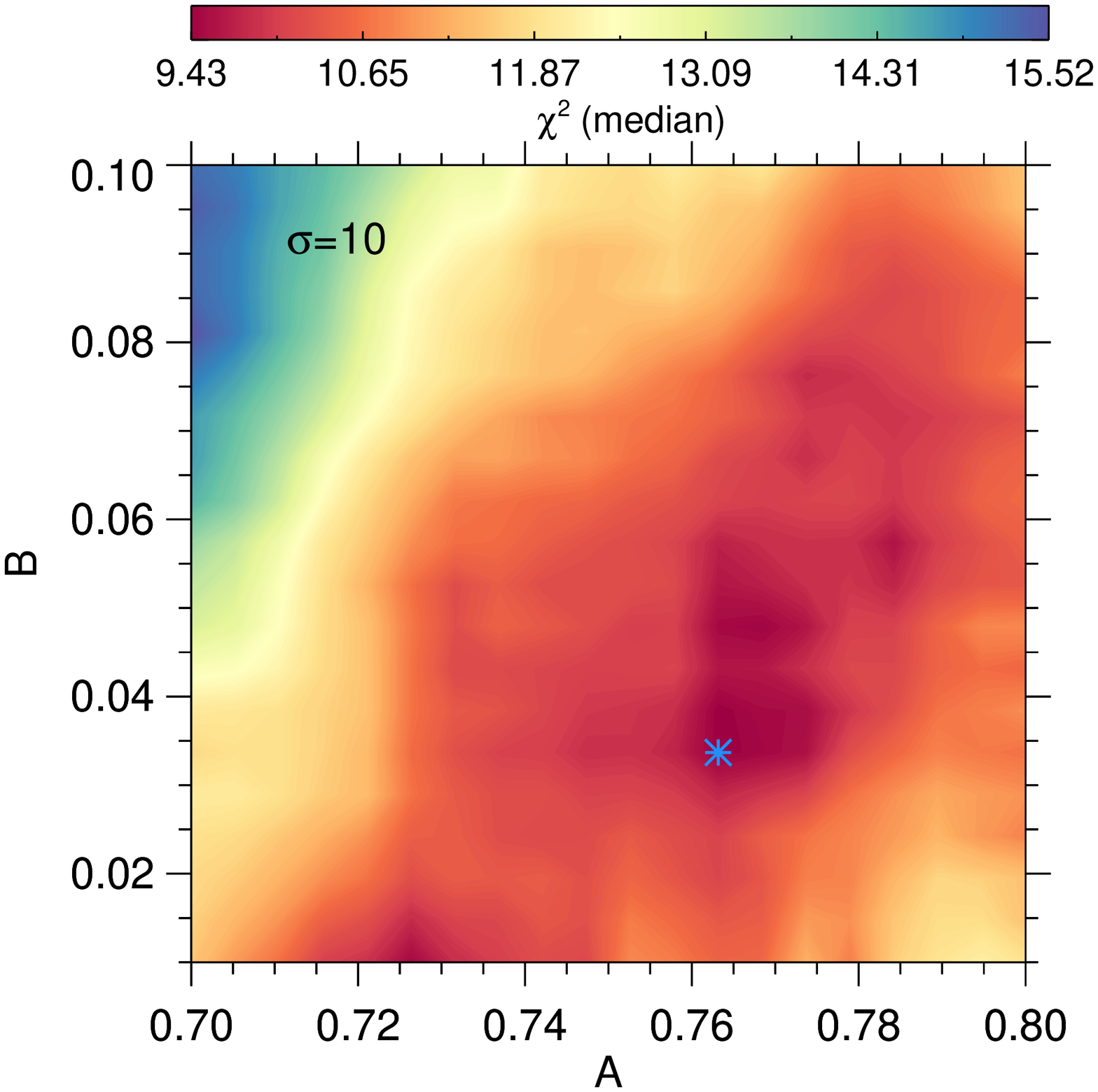} 
\includegraphics[width=0.4\textwidth]{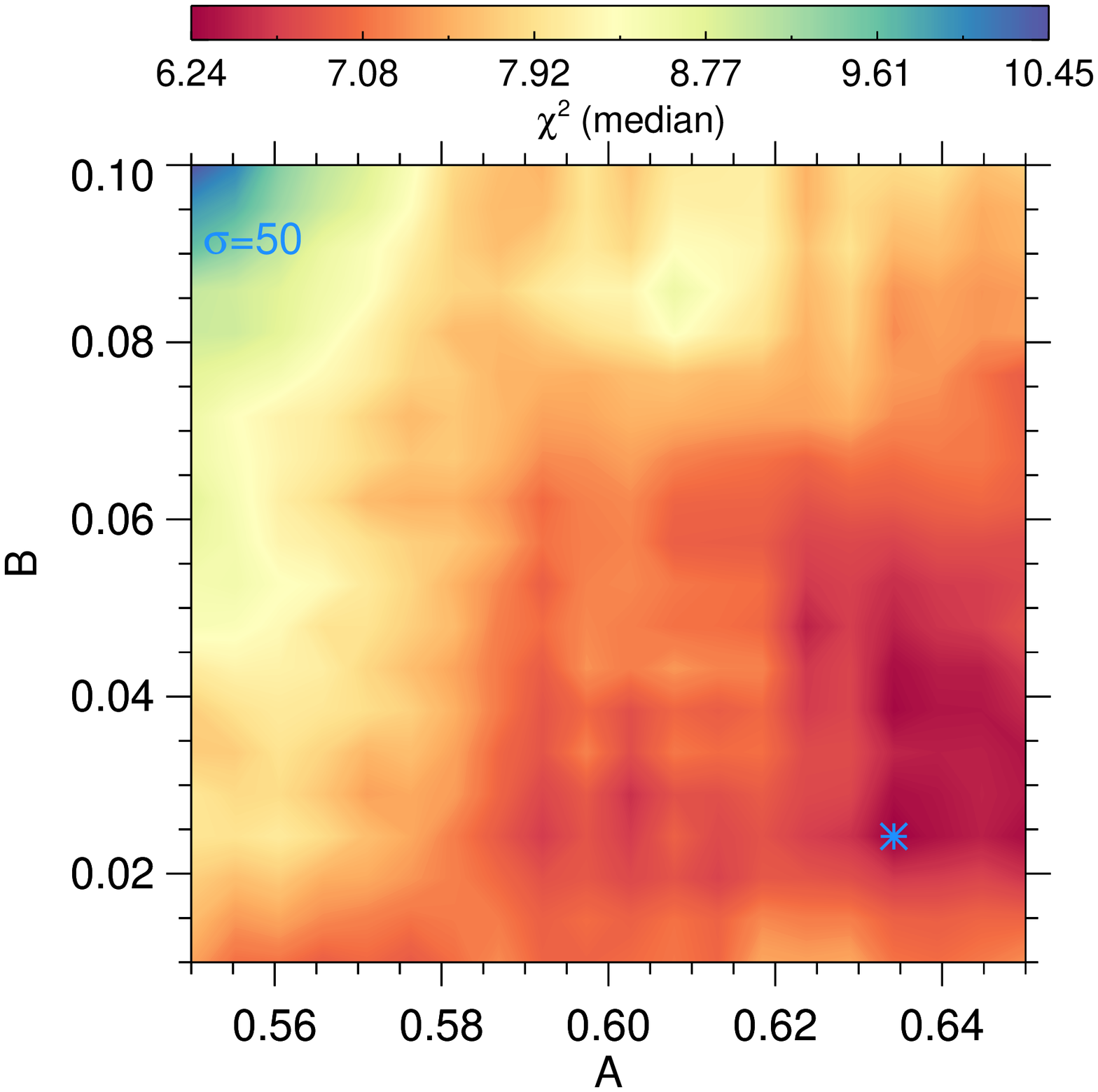} 
\caption{Maps of the median $\chi^2$ value obtained for 10 MC realizations for $\sigma=3$ to 50 (from top to bottom). 
The symbol marks the $A,B$ values that correspond to the minimum median $\chi^2$ value.}
\label{fig:maps}
\end{figure}

As a next step, we calibrate the suppression factor, namely we determine the free parameters $A,B$. To do so, we make a quantitative comparison of the plasmoid size distributions obtained from the PIC simulations of SGP16 and our MC code. For each plasma magnetization we:
\begin{enumerate}
 \item create a $20\times20$ grid of $A,B$ values. 
 \item run the MC code for every pair of $A,B$ values from the grid. 
 \item compute the {\sl frequency} histograms of $\log w$ for all plasmoids formed in our MC and PIC simulations using the same number of bins $N$.
 \item introduce a  measure of the goodness of fit. The test statistic is defined as:
 \eqb 
 \chi^2 = \sum_{i=1}^{N}\frac{\left(h_i^{\rm (MC)} - h_i^{\rm (PIC)} \right)^2}{h_i^{\rm (PIC)}},
 \label{eq:chi2}
 \eqe 
where $h_i^{\rm (PIC)}$ and $h_i^{\rm (MC)}$ are the values of the frequency histograms computed for the PIC and MC simulations, respectively. The sum is over the bins in $\log w$. We caution the reader that the absolute $\chi^2$ values calculated by eqn.~(\ref{eq:chi2}) should not be taken at face value, since systematic uncertainties in both experiments (PIC and MC) are poorly defined. The distribution of $\chi^2$ values for different choices of the $A,B$ parameters in the MC simulations is what will be used in our analysis. We also compared the distributions using a generalization of the classical $\chi^2$ test \citep[eqn.~5 in][]{gagunashvili_10} and we obtained the same qualitative results as those presented in Fig.~\ref{fig:maps}. 
 \item create a map of the $\chi^2$ values and repeat steps (ii)-(iv) for 10 different MC realizations. This is necessary, since the test statistic is subjected to the randomness of the MC and PIC simulations. 
 \item compute the median and the standard deviation of the $\chi^2$ values obtained from the multiple MC realization for every pair of $A,B$ values. The median $\chi^2$ map is expected to be smoother, since features related to by-chance bad fits can be canceled out. Regions of the $A-B$ parameter space with a large standard deviation in the $\chi^2$ values are more variable due to the randomness of the process.
 \item calibrate our MC code using the $A,B$ values that correspond to the minimum $\chi^2$ value of the median map. 
\end{enumerate}
Our results are presented in Fig.~\ref{fig:maps} for $\sigma=3, 10$, and 50 (from top to bottom). 
In all cases, we find an extended diagonal region of lower $\chi^2$ values. This indicates a correlation between $A$ and $B$, which can be understood as follows:  parameter $A$ sets a threshold in plasmoid momentum above which the suppression of the plasmoid growth becomes significant, while parameter $B$ determines the degree of the suppression; lower $B$ values lead to larger suppression factors. 
Although our MC code can reproduce satisfactorily the results of PIC simulations for $A,B$ values drawn from the deep red coloured regions of the maps, we will calibrate our code using the $A,B$ as indicated on the plot with blue stars (see Sect.~\ref{sec:results}).

\section[]{Analytical expressions for the plasmoid size distribution}
\label{app:app1} 
We wish to obtain an analytical solution for the steady-state plasmoid distribution $N(w)\equiv {\rm d}N/{\rm d}w$. The governing equation for the size- and position-dependent plasmoid distribution, $n(w, x; x_0)$,  is a partial differential equation  (PDE) of the general form:
	\eqb
	\label{eqn:n_kinetic_eqn_1}
	\frac{{\rm d} n}{ {\rm dt}} = \tilde{Q} + \tilde{L},
	\eqe
	where $\tilde{Q}$ and $\tilde{L}$ represent source and sink terms, respectively. The left-hand side of the eqn.  can be expanded as the following
	\eqb
	\label{eqn:dt_eqn}
	\frac{{\rm d} n}{ {\rm dt}} = v \, \partial_{x} n + \frac{{\rm d}w} {{\rm d}t} \, \partial_{w} n,
	\eqe
	where $\partial_{\rm t} n = 0$ in  steady state and  $v\equiv {\rm d} x/ {\rm d}t$ is the plasmoid velocity, which can be written as:
	\eqb
	\label{eqn:velocity_def}
	v(w, x; x_0) = \frac{p(w, x ; x_0) \, c}{\sqrt{1+ p(w, x ;x_0)^2}}.
	\eqe
	Here, $x_0$ is the plasmoid's birth location and $p \equiv \beta \gamma$ is its dimensionless  four-velocity (or, momentum) that also depends implicitly on time $t$ through $w$ and $x$  (see eqn.~(\ref{eq:p})). Equation (\ref{eqn:dt_eqn}) can be written as:
	\eqb
	\label{eqn:dt_eqn_2}
	\frac{{\rm d} n}{ {\rm d}t} = v \, \left(\partial_{x} n + \frac{\beta_{\rm g}}{p \,  f_{\rm sup}} \, \partial_{w}n\right),
	\eqe
	where we made use of ${\rm d} w / {\rm d}t = v \, ({\rm d}w/ {\rm d}x)$ and ${\rm d}w/ {\rm d}x = \bg / (p f_{\rm sup})$ \citep[see also eqn.~(4) in][]{pgs_16}.  
	The production rate of new plasmoids ($\tilde{Q}$) is given by:
	\eqb
	\label{eqn:injection_term}
	\tilde{Q} = q_{0} \, \delta(w-w_*) \, S(x; 0, L),
	\eqe
	where $q_0$ is a normalization constant and $S(y; y_1, y_2)$ is the unit box function. The source term as defined above describes a uniform generation of plasmoids with size $w_*$ across the layer. The loss of plasmoids occurs either by the advection from the reconnection layer ($\tilde{L}_{\rm esc}$) or through the coalescence of two plasmoids ($\tilde{L}_{\rm m}$). The advection term is defined as $\tilde{L}_{\rm esc} = - n/t_{\rm esc}$, where the escape time is defined as 
	\eqb
	\label{eqn:escape_time}
	t_{\rm esc}(w, x; x_0) = \epsilon \int_{x}^{L} \frac{{\rm d} y}{v(w, y ;x_0)}, \, \epsilon \ge 1.
	\eqe
	A choice of  $\epsilon \gg 1$ ensures that $t_{\rm esc} \gg t_{\rm A}\simeq (L-x)/v_{\rm A}$. Thus, plasmoids leave the layer only when they are located close to the edge, as it is observed in PIC simulations.   Replacing the injection and advection terms with their explicit expressions, eqn. ~(\ref{eqn:n_kinetic_eqn_1}) becomes
	\eqb
	\label{eqn:n_kinetic_eqn_2}
	\partial_{x} n + \frac{\bg  \partial_{w} n}{p \, f_{\rm sup}} = \frac{q_{0}\, \delta(w-w_*) \, S(x; 0, L)}{v} - \frac{n}{v \, t_{\rm esc}}+\tilde{L}_{\rm m}.
	\eqe
	
Equation~\ref{eqn:n_kinetic_eqn_2} can be solved by the method of characteristics to obtain a family of solutions $n(w,x;x_0)$ for different values of $x_0$. The plasmoid distribution $N(w,x)$ can be then expressed as the weighted average of $n(w,x;x_0)$ over $x_0$, namely
\eqb
\label{eqn:Nwx_def}
N(w,x) = \frac{\int_0^L {\rm d}x_0 n(w,x;x_0) z(x_0)}{\int_0^L {\rm d}x_0 z(x_0)},
\eqe
where $z(x_0)$ is a weighting function; in the simplest scenario, $z(x_0)=1$.  The differential size distribution is then obtained as  $N(w) = \int {\rm d}x N(w,x)$. 

The first characteristic equation of the PDE is 
\eqb 
\label{eqn:dwdx_1}
\frac{{\rm d}w}{{\rm d}x} = \frac{\beta_{\rm g}}{p(w,x; x_0) f_{\rm sup}(|p|)}.
\eqe 
This can be solved only numerically in the most general case where $p$ and $f_{\rm sup} \ne 1$ are given respectively by eqns~(\ref{eq:p}) and  (\ref{eq:sup}). 

We thus derive analytical solutions of eqn.~(\ref{eqn:n_kinetic_eqn_2}) under the following simplifying assumptions: 
\begin{enumerate}
         \item the growth rate is constant (e.g., $\bg=0.1\beta_{\rm g,-1}$) and there is no growth suppression  ($f_{\rm sup}=1$). 
         \item plasmoids are lost from the layer only due to advection, i.e., $\tilde{L}_{\rm m}=0$.
         \item the plasmoid momentum is constant and equal to its asymptotic value, i.e., $p=\sqrt{\sigma} > 1$, or
         the plasmoid momentum is non-relativistic and increases linearly with $x/w$, namely $p \approx \bacc (x-x_0)/w$ (see eqn.  (\ref{eq:p})). 
\end{enumerate}
\subsection{Plasmoids with terminal momentum} 
The solution to the characteristic eqn.~(\ref{eqn:dwdx_1}) is $w-w^\prime =A_{\rm g}(x-x^\prime)$, 
where $A_{\rm g}=\bg/\sqrt{\sigma}$ and $w^\prime, x^\prime$ are arbitrary initial values. The solution to eqn.~(\ref{eqn:n_kinetic_eqn_2}) is then calculated as 
\eqb
n(w,x;x_0) = \frac{q_0}{v_A A_{\rm g}}\int_{w_1}^{w_2}{\rm d}w^\prime \delta(w^\prime-w_*)\left(\frac{L-x}{L-x+\frac{w-w^\prime}{A_{\rm g}}}\right)^{1/\epsilon} 
\eqe
where $w_1=w-A_{\rm g} x$ and $w_2=w-A_{\rm g} (x-L)$. For $w_1< w_* < w_2$ or, equivalently $w_*-A_{\rm g}(L-x) < w <w_* + A_{\rm g} x$, the above equation results in
\eqb
\label{eqn:Nwx_sol1}
N(w,x) =\frac{q_0}{v_A A_{\rm g}}\left(\frac{L-x}{L-x+\frac{w-w_*}{A_{\rm g}}}\right)^{1/\epsilon}.
\eqe
At $x\ll L$ most of the plasmoids have $w\gtrsim w_*$, whereas  plasmoids with larger sizes exist at larger distances from the center of the layer. The reason is that an individual plasmoid grows in size with a constant rate $\beta_{\rm g}$ as it moves along the layer $x$. The maximum size of the plasmoid chain is 
\eqb 
 \wmx =w_* + A_{\rm g} L.
\label{eqn:wmax}
\eqe 

We find the differential size distribution by integrating $N(w,x) $along the layer:
\eqb
N(w) & = & \frac{q_0}{v_A A_{\rm g}} \int_{x_1}^{x_2} {\rm d}x \left(\frac{L-x}{L-x+\frac{w-w_*}{A_{\rm g}}}\right)^{1/\epsilon} \\ 
N(w) &  \xrightarrow{\epsilon \gg 1} & \frac{Q_0}{A_{\rm g}}\left(L -\frac{w-w_*}{A_{\rm g}} \right),
\label{eqn:Nw_sol1}
\eqe
where $Q_0=q_0/v_A$, $x_1=\max[0, (w-w_*)/A_{\rm g}]=(w-w_*)/A_{\rm g}$ and $x_2=\min[L, L+(w-w_*)/A_{\rm g}]=L$. For $\epsilon \gg 1$ plasmoids can leave the layer only when they are located close to the edge, as it is observed in PIC simulations. We find that $N(w) \propto$ const for $w \ll \wmx$, while it decreases fast as $w \rightarrow \wmx$.  
\subsection{Accelerating plasmoids} 
Let us solve eqn. ~(\ref{eqn:n_kinetic_eqn_2}) in the limit where 
\eqb 
p \approx \frac{\beta_{\rm a} (x-x_0)}{w}
\label{eqn:f_approx}
\eqe 
and $p_0 \approx 0$. This approximation  is valid when following condition is satisfied:
\eqb
\frac{\beta_{\rm a} (x-x_0)}{w \sqrt{\sigma}} <1.
\label{eqn:f_approx_cond}
\eqe
The characteristic equation  of the PDE now reads ${\rm d}w/{\rm d}x=\bg w/[\bacc (x-x_0)]$ and its solution is given by 
\eqb
\label{eqn:dwdx_sol2}
\frac{w}{w^\prime} = \left(\frac{|x-x_0|}{|x^\prime -x_0|}\right)^{s}
\eqe
where
\eqb
s\equiv \frac{\beta_{\rm g}}{\beta_{\rm a}}.
\label{eqn:s}
\eqe 
The solution to the eqn.~(\ref{eqn:n_kinetic_eqn_2}) can be then calculated as 
\eqb
\label{eqn:Nwx_sol2}
n(x,w;x_0)& = & q_0\int_0^{L}{\rm d}x^\prime \frac{\delta(w^\prime-w_*)}{v(x^\prime, w^\prime; x_0)} \\ \nonumber
& = & \frac{q_0 \beta_{\rm a}}{\beta_{\rm g}} \int_{w_1}^{w_2}{\rm d}w^\prime \frac{ (x^\prime-x_0)\delta(w^\prime-w_*)}{w^\prime v(x^\prime, w^\prime; x_0)},
\eqe
where the integration limits are: 
\eqb
w_1 & = & w \left(\frac{x_0}{x-x_0}\right)^s \\
w_2 & = & w \left(\frac{L-x_0}{x-x_0}\right)^s.
\eqe
Substitution of eqn. ~(\ref{eqn:dwdx_sol2}) to eqn. ~(\ref{eqn:Nwx_sol2}) and performance of the integral leads to 
\eqb
\label{eqn:nwx_sol3}
n(w,x; x_0)= \frac{q_0}{c\beta_{\rm g}}\sqrt{1+\left(\frac{\beta_{\rm a}(x-x_0)}{w_*}\left(\frac{w_*}{w}\right)^{1/s}\right)^2},
\eqe 
for plasmoid sizes satisfying the following conditions:
\eqb
w_*\left(\frac{x-x_0}{L-x_0}\right)^s < w < w_*\left(\frac{x-x_0}{x_0}\right)^s \\
w > \frac{\beta_{\rm a}(x-x_0)}{\sqrt{\sigma}}.
\eqe
In the limit where the approximation of momentum is valid (see eqn. ~(\ref{eqn:f_approx_cond})) and for plasmoids with $w/w_* > \sqrt{\sigma}^{s/(1-s)}$, the square root in eqn.~(\ref{eqn:nwx_sol3}) can be approximated by $\sqrt{1+u^2}\approx 1+u^2/2$. Integration of eqn.~(\ref{eqn:nwx_sol3}) with respect to $x$ results in:
\eqb
\label{eqn:nw_sol3_x0}
n(w;x_0) = \frac{q_0}{\beta_{\rm g} c}\left[x+\frac{\beta_{\rm a}^2(x-x_0)^3}{6 w_*^2}\left(\frac{w_*}{w}\right)^{2/s}\right]_{x_1}^{x_2},
\eqe
where $x_1 = x_0+x_0 (w/w_*)^{1/s}$, $x_2=\min(L, x_0+ w/\wbr)$, and $\wbr \equiv \bacc L / \sqrt{\sigma}$. This is the typical size of plasmoids that accelerate and exit the layer with $p\approx \sqrt{\sigma}$, while even larger plasmoids are, in general, slower \citep[see also][]{pgs_16}.

The upper integration limit is $x_2=L$ when $x_0 > L-\frac{\sqrt{\sigma} w}{\beta_{\rm a}} = L (1-w/\wbr)$. Additionally, $x_2 > x_1$ only if $x_0 < L/[1+\zeta^{1/s}]$ where $\zeta=w/w_*$. The lower and upper limits for integration over $x_0$ (see  eqn.~(\ref{eqn:nw_sol3_x0})) are respectively
$x_{0,1} = \max[0, L(1-w/\wbr)]$ and $x_{0,2}=L/[1+\zeta^{1/s}]$. We are interested in the plasmoid distribution of plasmoids with $w \gtrsim \wbr$. For $w> \wbr$, the plasmoid distribution is given by:
 \eqb 
 N(w) = \frac{q_0 L}{2\beta_{\rm g} c} \left[\frac{1}{1+\zeta^{1/s}}+\frac{\beta_{\rm a}^2 L^2}{12 w_*^2 \zeta^{2/s}}\left(1-\frac{\zeta^{3/s}}{\left(1+\zeta^{1/s}\right)^3} \right)
  \right]
 \eqe 
 which for $w \gg w_*$ simplifies in:
 \eqb
  \label{eqn:nw_sol3}
  N(w) \approx \frac{q_0 L}{2\beta_{\rm g}c}\zeta^{-1/s}. 
  \eqe
 Thus, the differential size distribution of plasmoids with a position- and size-dependent momentum which is $\lesssim \sqrt{\sigma}$ can be described by a power law with slope $-\beta_{\rm a}/\beta_{\rm g}$. For plasmoids with $w > \wbr$ mergers are less frequent compared to the smaller plasmoids  (see Fig.~\ref{fig:merging-2}). It is therefore safe to ignore the merger term in eqn.~(\ref{eqn:n_kinetic_eqn_2}). Furthermore, their growth rate is not suppressed, for they do not move with the terminal momentum. 
\end{document}